\documentclass{article}
\usepackage[utf8]{inputenc}
\usepackage{amsfonts, amsmath, amssymb, amsthm,bm,bbm}
\usepackage[margin=0.8in]{geometry}
\usepackage{xcolor}
\usepackage{diagbox}
\usepackage{tabu}
\usepackage{multirow}
\usepackage[pagebackref, colorlinks = true, linkcolor = blue, urlcolor  = brown, citecolor = red]{hyperref}
\usepackage{tikz}
\usepackage{authblk}
\newcommand\diag[4]{%
  \multicolumn{1}{p{#2}|}{\hskip-\tabcolsep
  $\vcenter{\begin{tikzpicture}[baseline=0,anchor=south west,inner sep=#1]
  \path[use as bounding box] (0,0) rectangle (#2+2\tabcolsep,\baselineskip);
  \node[minimum width={#2+2\tabcolsep-\pgflinewidth},
        minimum  height=\baselineskip+\extrarowheight-\pgflinewidth] (box) {};
  \draw[line cap=round] (box.north west) -- (box.south east);
  \node[anchor=south west] at (box.south west) {#3};
  \node[anchor=north east] at (box.north east) {#4};
 \end{tikzpicture}}$\hskip-\tabcolsep}}

\newcommand{\Tr}{\mathrm{Tr}}
\newcommand{\bE}{\mathbb{E}}
\newcommand{\bM}{\mathbb M}
\newcommand{\cM}{\mathcal{M}}
\newcommand{\cW}{\mathcal{W}}
\newcommand{\s}{s}
\newcommand{\RNum}[1]{\uppercase\expandafter{\romannumeral #1\relax}}

\newtheorem{definition}{Definition}

\newtheorem{remark}{Remark}

\title{Schwinger-Dyson and loop equations for a product of square Ginibre random matrices}
\author[1]{{Stephane~Dartois}} 
\author[1,2]{{Peter~J.~Forrester}}
\affil[1]{School of Mathematics and Statistics, and  
${}^2$ARC Centre of Excellence for Mathematical \& Statistical Frontiers,
University of Melbourne, Victoria 3010, Australia\\
    stephane.dartois@unimelb.edu.au and pjforr@unimelb.edu.au}
    
\date{\today}
\begin{document}
\maketitle
\begin{abstract}
    In this paper, we study the product of two complex Ginibre matrices and the loop equations satisfied by their resolvents (\textit{i.e.}~the Stieltjes transform of the correlation functions). We obtain using Schwinger-Dyson equation (SDE) techniques the general loop equations satisfied by the resolvents. In order to deal with the product structure of the random matrix of interest, we consider SDEs involving the integral of higher derivatives. One of the advantage of this technique is that it bypasses the reformulation of the problem in terms of singular values. As a byproduct of this study we obtain the large $N$ limit of the Stieltjes transform of the $2$-point correlation function, as well as the first correction to the Stieltjes transform of the density, giving us access to corrections to the smoothed density. In order to pave the way for the establishment of a topological recursion formula we also study the geometry of the corresponding spectral curve. This paper also contains explicit results for different resolvents and their corrections.
\end{abstract}

\vspace{3cm}
\vspace{2cm}

\section{Introduction}

The study of random matrices in mathematics can be traced back to the work of Hurwitz on the invariant measure for the matrix groups $U(N)$ and $SO(N)$ \cite{Hu97,DF17}. In multivariate statistics another stream of random matrix theory was initiated 
with the work of Wishart \cite{wishart1928generalised} on estimating the covariance matrices of multivariate statistics when the number of variables is large. In theoretical physics Wigner \cite{wigner1955characteristic} used random matrices to model energy spectrum of Hamiltonians of highly excited states of heavy nuclei. The works of physicists \cite{t1974planar} on the large $N$ limit of $U(N)$ gauge theory provided yet another application to random matrices (and their generalized version often referred to as \emph{matrix models}). Since then random matrix theory and matrix models have been found useful in an overwhelming number of contemporary fields, for example communication engineering \cite{TV04}, the analysis of algorithms \cite{Tr15}, and deep learning \cite{PW17}.
 Many tools have been developed to understand the properties of different models and ensembles. One of these tools is called loop equations, and has led to the now well-known Chekhov-Eynard-Orantin topological recursion formula \cite{eynard2004all, chekhov2006free, chekhov2006hermitian}. In the realm of random matrix theory this formula allows for the  systematic computation of correlation functions of random matrices, as series in $1/N$. \\

However some random matrix ensembles are, in the existing literature, still out of the scope of these loop equations. These are product ensembles, that is they are random matrices constructed out of a product of several random matrices. In this paper we describe the loop equations for such a product ensemble, specifically considering the case of a random matrix constructed out of the product of two complex Ginibre matrices. Such an ensemble was for instance considered in \cite{Bouchaud2007}, with applications to the study of financial data, while a closely related product ensemble with applications to low energy QCD, was studied in \cite{Osborne-QCD-2-matrices} (see also the text book treatment
\cite[\S 15.11]{Fo10}), allowing for insight into the poorly understood regime of non-zero baryon chemical potential. \\

More generally the product ensembles are found to have many applications. Some of these applications are described in the thesis \cite{Jesper-thesis}. Among those, one finds applications to telecommunication problems where product ensembles provide a model of communication channels where the signal has to pass through different media \cite{muller-telecom-channel-2002}. One also finds applications to the study of spin chains with disorder \cite{prod-rand-mat-book}, quantum transport \cite{Beenakker-review-Qtransport}, quantum information and random graph states \cite{collins2010randoma, collins2013area}. The product ensembles also relate to the study of neural networks. Indeed information about the asymptotic behavior of such ensembles allows one to draw results about stability of gradient in a deep neural network with randomly initialized layers \cite{hanin2018products}. These product ensembles are also of interest for the study of the stability of large dynamical systems \cite{BEN-dyn-syst-RMT,IF18}. As a consequence, finding mathematical and technical tools for investigating the properties of these ensembles can enable progress in these fields of study.  \\

Yet another problem of importance is the one of Muttalib-Borodin ensembles. These ensembles were first defined as invariant ensembles, via their eigenvalue probability density function (PDF) \cite{muttalib1995random}, and latter realized in terms of ensembles of random matrices with independent entries \cite{Ch14,FW15}. Their joint PDF is proportional to,
\begin{equation}\label{A}
    \prod_{l=1}^N e^{- V(\lambda_l)} \prod_{1 \le i < j \le N} (\lambda_i - \lambda_j)(\lambda_i^\theta - \lambda_j^\theta), 
\end{equation}
where $\theta > 0$ is a parameter and $V(\lambda_l)$ can be interpreted as a confining potential. For general potential $V$ and $\theta=2$, this model relates to the $\mathcal{O}(\mathfrak{n})$ matrix model with $\mathfrak{n}=-2$, see \cite{Borot-Eynard-O(n)}, and it also relates to a particular model of disordered bosons \cite{LSZ06}. A key structural interest in the  Muttalib-Borodin ensembles is that they are biorthogonal ensembles. That is they admit a family of biorthogonal polynomials and their correlation functions can be expressed in determinantal form, with a kernel that can be expressed in terms of the biorthogonal polynomials; see \cite{borodin1998biorthogonal}. Although it is not immediately obvious, the singular values for the product of $M$ complex Ginibre matrices also give rise to biorthogonal ensembles \cite{AIK13,KZ14}. Moreover, in the asymptotic regime of large separation, the PDF for the squared singular values reduces to (\ref{A}) with $\theta = 1/M$, and $V$ having the leading form $V(x) = - M x^{1/M}$ \cite{FLZ15}.\\

One attractive feature of both the Muttalib-Borodin ensemble, and the squared singular values of products of complex Ginibre matrices, is that in the global density limit the moments of spectral density are given by the Fuss-Catalan family of combinatorial numbers; see 
\cite{penson2011product,FW15}. Another is the special role played by particular special functions of the Meijer-G and Wright Bessel function class. Underlying these special functions is a linear differential equation of degree $M +1$. Less well understood is the nonlinear differential system implied by the correlation kernel based on these special functions. These are relevant to the study of gap probabilities; see \cite{WF17,MF18}. \\

Other questions about products of random matrices have been investigated for instance in \cite{Dubach-Peled}. In this work, the authors are concerned about the behavior of traces of general words of Ginibre matrices. In particular they show that the limiting square singular values distribution is a Fuss-Catalan distribution for any words. In the work \cite{DLN}, the authors study the traces of the general words in an alphabet of random matrices constructed out of the marginals of a random tensor. Using combinatorial techniques it is possible to show freeness of some marginals or to describe entirely the free cumulants when there is no freeness of the different marginals in the limit. One interesting aspect is that using these products of marginals it is possible to find distribution interpolating between the square of a Mar\u cenko-Pastur law and the free multiplicative square of a Mar\u cenko-Pastur law.\\

However there is in general little technical tools to describe the lower order in $N$ observables of product ensembles. Indeed free probability provides us with some useful techniques (free additive and multiplicative convolution), but those are restricted to the large $N$ limit, and comes in handy only for the study of the large $N$ density or the behavior of the large $N$ limit of the moments (with some extension to the fluctuations of the linear statistics \textit{via} \cite{collins2007second}).\\

In this paper we focus on describing the loop equations for the random matrix $S_2=X_1X_1^{\dagger}X_2^{\dagger}X_2$, where $X_1, X_2$ are square complex Ginibre matrices. In order to obtain these loop equations we start with Schwinger-Dyson identities and use them to obtain relations between moments, later translated in terms of equations on the resolvents of $S_2$. These equations on the resolvents are the loop equations. One of the new features of the method presented here is that the starting point Schwinger-Dyson identities involve higher order derivatives. This allows us to obtain relations between moments of the matrix $S_2$ only without having to deal with mixed quantities.
Thanks to the combinatorial interpretation of the moments of the matrix $S_2$ (that we also shortly describe), we show that the (connected) resolvents possess a $1/N$ expansion, which is the unique additional ingredient we need to be able to solve the loop equations recursively. \\

Using this data we illustrate the use of the obtained loop equations by computing the large $N$ limit of the resolvent $W_{0,1}(x)$, thus recovering known results relating to the generating function of the moments. We also compute $W_{0,2}(x_1,x_2)$ (that is the Stieltjes transform of the $2$-point correlation function) and show that it takes the expected universal form once expressed in the correct variables, thus relating to the Bergmann kernel on the sphere. We give explicit results for $W_{1,1}(x), W_{2,1}(x)$ (first and second correction to the large $N$ limit of the resolvent), $W_{1,2}(x_1,x_2)$ (first correction to $W_{0,2}(x_1,x_2)$), as well as $W_{0,3}(x_1,x_2,x_3)$. One interesting aspect of the obtained loop equations are their structural properties, that seem to generalize in a very natural way the usual bilinear loop equations for random matrices or matrix models. In particular, the family of loop equations we obtain for this product of matrices are trilinear in the resolvents $W_{g,n}$. This is at the root of the appearance of the double ramification point of $W_{0,1}(x)$ and we expect that a topological recursion formula similar to the one obtained in \cite{Bouchard-Eynard} applies. Moreover they contain generalizations of the derivative difference term usually appearing in the bilinear setting, as well as derivatives of first and second order. Motivated by these interesting structural properties, we use the explicit computations to explore the analytical properties of the $W_{g,n}$ (or rather their analytic continuation on the associated spectral curve). These explorations give further hint that there is a topological recursion formula to compute them systematically. We expect that a similar technique allows to describe the loop equations for the product of $p\ge 2$ rectangular Ginibre matrices $S_p=X_1X_2\ldots X_p(X_1X_2\ldots X_p)^{\dagger}$; we leave this study, as well as the one of a topological recursion formula, to further works. Note that, as a byproduct, we also expect that this technique applies to the interesting matrix models introduced in \cite{A-C2014,A-C2018} to generate hypergeometric Hurwitz numbers.

\paragraph{Organisation of the paper.}
The paper is organized as follows. In section \ref{sec:Wishart}, we use the Wishart case (that is the case of one Ginibre matrix) as a pedagogical example. It is used to sketch the combinatorial arguments allowing to show the existence of the $1/N$ expansion and to illustrate the Schwinger-Dyson equation technique in a simpler context. The reader already accustomed to Schwinger-Dyson equations obtained using the matrix elements variables and knowledgeable on the associated combinatorics may consider skipping this section.

In section \ref{sec:loop-eq-prod}, we describe the heart of this paper, that is the derivation of the Schwinger-Dyson equations and loop equations for a product matrix of the form $S_2=X_1X_1^{\dagger}X_2^{\dagger}X_2$. The loop equations take the form of a family of equations on the resolvents, that is the Stieltjes transforms (denoted $W_n(x_1,\ldots,x_n)$) of the $n$-point correlation functions. We present the results step by step to make the method transparent to the reader and the first few special cases that are the loop equations for $W_1(x)$, $W_2(x_1,x_2)$ and $W_3(x_1,x_2,x_3)$ are presented in details. This section ends with the main result, that is the loop equations satisfied by any $W_{g,n}(x_1,\ldots,x_n)$ as shown on equation \eqref{eq:loop-eq-general-expanded}, where $W_{g,n}(x_1,\ldots,x_n)$ is the coefficient of order $g$ of the $1/N$ expansion of $W_n(x_1,\ldots,x_n)$.

In section \ref{sec:spectral-curve-geometry}, we take on a geometrical point of view in order to compute the $W_{g,n}$ more effectively from the loop equations. We describe in details the \emph{spectral curve} geometry associated to the problem. We compute after a change of variables, $W_{0,2}(x_1,x_2)$, $W_{1,1}(x)$, $W_{2,1}(x)$, $W_{1,2}(x_1,x_2)$ and $W_{0,3}(x_1,x_2,x_3)$ (see equations \eqref{eq:W02}, \eqref{eq:W11}, \eqref{eq:W21}, \eqref{eq:W12}, \eqref{eq:W03}). We use these explicit computations to explore the analytic properties of the loop equations. These properties are expected to be of importance to establish a topological recursion formula allowing to systematically compute every $W_{g,n}$.

\section*{Acknowledgments}
Stephane Dartois would like to thank Valentin Bonzom, Alexandr Garbali, Jesper Ipsen and Paul Zinn-Justin for useful discussions and technical help related to this work as well as for references. This work was supported by the Australian Research Council grant DP170102028.

\section{One matrix case, Wishart ensemble}\label{sec:Wishart}
In this section, we illustrate the problem that is our interest in this paper on a simpler case, that is the (trivial) product of one matrix. This is the case of a Wishart matrix. We first recall the combinatorial representation of moments of a Wishart ensemble matrix. We then show how we can compute the average resolvent of a Wishart matrix using the Schwinger-Dyson equation method. It is only in the next section that we consider the case of the product of two Ginibre matrices. Thus the technically knowledgeable reader can skip this section and start reading section \ref{sec:loop-eq-prod}.
\subsection{Random Wishart matrices}
In this paper we always consider square matrices. In the Wishart matrices case it corresponds to setting the asymptotic size ratio parameter $c$ to $1$.
Let $X\in \mathcal{M}_{N\times N}(\mathbb{C})$ be a Ginibre random matrix. More concretely, $X$ is a random matrix whose entries are i.i.d. complex Gaussian with zero mean, or more formally, the entries $X_{i,j}$ are distributed according to the  density
\begin{equation}
    \frac{N}{2i\pi}e^{-N\lvert X_{i,j}\rvert^2}\mathrm{d}\bar{X}_{i,j}\mathrm{d}X_{i,j}.
\end{equation}
In particular we denote,
\begin{equation}
    \mathrm{d}X^{\dagger}\mathrm{d}X=\prod_{i,j}\mathrm{d}\bar{X}_{i,j}\mathrm{d}X_{i,j},
\end{equation}
so that $X$ has the  distribution
\begin{equation}
    \mathrm{d}\mu(X)=\frac{N^{N^2}}{(2i\pi)^{N^2}}e^{-N\Tr(XX^{\dagger})}\mathrm{d}X^{\dagger}\mathrm{d}X.
\end{equation}
A (complex) Wishart random matrix is the random variable defined as the product $S_1=XX^{\dagger}$.\\
\

\begin{figure}
    \centering
    \includegraphics[scale=0.8]{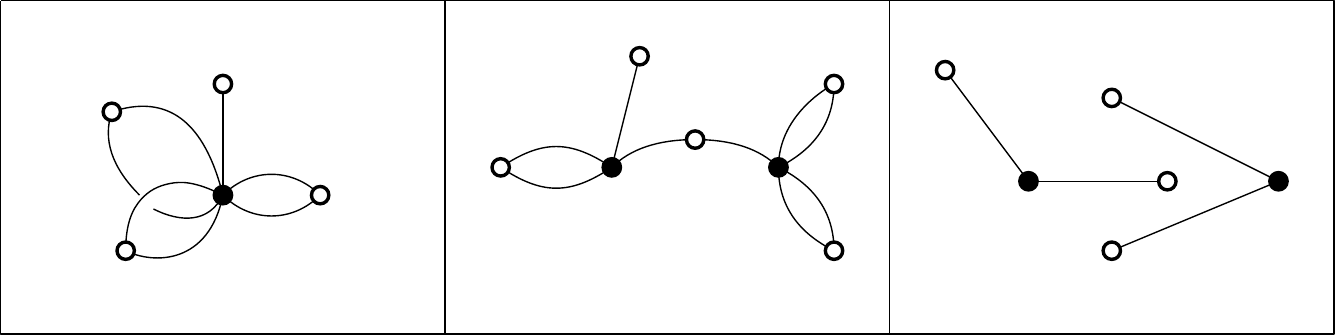}
    \caption{Left: Map of genus $1$ contributing to the computation of $m_7$. Center: Connected map of genus $0$ contributing to the computation of $c_{4,5}$ and also to $m_{4,5}$. Left: Disconnected map with two genus $0$ components. Contribute to the computation of $m_{2,2}$. }
    \label{fig:map_examples_Wishart}
\end{figure}

\noindent{\bf Combinatorics of moments.} The moments $m_k$ of order $k$ of a Wishart random matrix are defined as
\begin{equation}
    m_k=\bE\left(\Tr(S_1^k)\right).
\end{equation}
Further, for any sequence of positive integers $k_1,\ldots,k_n$ we can define moments $m_{k_1,\ldots,k_n}$ of order $k_1,\ldots,k_n$. Similarly to the moments of order $k$ they are defined as the expectation of products of traces of powers of $S_1$
\begin{equation}
    m_{k_1,\ldots,k_n}=\bE\left(\prod_{i=1}^n\Tr(S_1^{k_i})\right).
\end{equation}
As is for instance explained in \cite{DLN}, the moments of order $k$ can be computed as a sum over labeled bicolored combinatorial maps $\cM$ with one black vertex. This combinatorial representation of moments implies that the moments have a $1/N$ expansion. That is
\begin{equation}\label{eq:genus-exp-moments}
    m_k=\sum_{g\ge 0}N^{1-2g}m_k^{[g]},
\end{equation}
where $m_k^{[g]}$ are the coefficients of this expansion. This is a crucial point that allows one to solve the loop equations recursively. Note also that this expansion is finite, that is here $g<k/2$. Let us be a bit more explicit on this point.\\
We recall the definition of labeled bicolored combinatorial maps with possibly more than one black vertex.
\begin{definition}
A labeled bicolored combinatorial map is a triplet $\cM=(E,\sigma_{\bullet},\sigma_{\circ})$ where,
\begin{itemize}
    \item $E$ is the set of edges of $\cM$
    \item $\sigma_{\bullet},\sigma_{\circ}$ are permutations on $E$
    \item $\cM$ is said to be connected if and only if the group $\langle \sigma_{\bullet}, \sigma_{\circ}\rangle$ acts transitively on $E$.
\end{itemize}
\end{definition}
The cycles of $\sigma_{\circ}$ are called white vertices, the cycles of $\sigma_{\bullet}$ are called black vertices, and the cycles of $\sigma_{\bullet}\sigma_{\circ}$ are called faces. Combinatorial maps can be represented graphically \cite{DLN, countingsurfaces} as they encode embeddings of graphs on surfaces. We give a few examples in Fig. \ref{fig:map_examples_Wishart}.
\\

We define the set of combinatorial maps $\bM_p=\{\cM=(E,\sigma_{\bullet},\sigma_{\circ})\mid E=\{1,\ldots,p\}, \sigma_{\bullet}=\gamma=(123\ldots p)\}$. One shows, using Wick-Isserlis theorem \cite{WickThm,isserlis1918formula}, that the moments of order $k$ can be written as a sum over combinatorial maps $\cM\in \bM_p$ (see \cite{DLN} for details)
\begin{equation}
    m_k=\sum_{\cM\in \bM_k}N^{V_{\circ}(\cM)-k+F(\cM)},
\end{equation}
where $V_{\circ}(\cM)$ is the number of white vertices of $\cM$ and $F(\cM)$ is the number of faces of $\cM$. Using the fact that $V_{\bullet}+V_{\circ}(\cM)-k+F(\cM)=2-2g(\cM)$, where $g(\cM)$ is the genus of the combinatorial map (that is the genus of the surface in which the corresponding graph embedds), one can show equation \eqref{eq:genus-exp-moments}. 
\vspace{2mm}
\begin{remark}
Note that elements of $\bM_p$ are necessarily connected as $\gamma$ acts transitively of $\{1,\ldots,p\}$.
\end{remark}
\vspace{2mm}
We now define the relevant set of maps for studying the moments of order $k_1,\ldots, k_n$. In this case we denote $p=\sum_{i=1}^n k_i$, $E=\{1,\ldots,p\}$ and $\gamma_{k_1,\ldots, k_n}=(12\ldots k_1)(k_1+1\ldots k_2)\ldots(k_{n-1}+1\ldots k_n)$
\begin{equation}
    \bM_{k_1,\ldots,k_n}=\{\cM=(E,\sigma_{\bullet},\sigma_{\circ})\mid  \sigma_{\bullet}=\gamma_{k_1,\ldots, k_n}\}.
\end{equation}
The maps in $\bM_{k_1,\ldots,k_n}$ are possibly non-connected as $\gamma_{k_1,\ldots, k_n}$ does not act transitively on the set of edges. Consequently we define the corresponding set of connected maps
\begin{equation}
    \bM_{k_1,\ldots,k_n}^c=\{\cM=(E,\sigma_{\bullet},\sigma_{\circ})\mid  \sigma_{\bullet}=\gamma_{k_1,\ldots, k_n}, \langle \sigma_{\bullet},\sigma_{\circ}\rangle \textrm{ acts transitively on } E\}.
\end{equation}
We state without proof\footnote{The proof is very similar to the one black vertex case, already appearing in \cite{DLN}.} that 
\begin{equation}
    m_{k_1,\ldots,k_n}=\sum_{\cM\in \bM_{k_1,\ldots,k_n}} N^{V_{\circ}(\cM)-p+F(\cM)},
\end{equation}
where $p=\sum_i k_i$. We can define the associated cumulants $c_{k_1,\ldots,k_n}$ of the moments, through their relation to moments
\begin{align}\label{eq:mom-cum-for-traces}
    m_{k_1,\ldots,k_n}=\sum_{K\vdash \{k_1,\ldots,k_n\}}\prod_{\kappa_i\in K}c_{\kappa_i}.
\end{align}
This relation is just the moment-cumulant relation for the family of random variables $\bigl\{R_{k_i}:=\Tr(S_1^{k_i})\bigr\}$.
These cumulants can be expressed as sums over connected combinatorial maps
\begin{equation}\label{eq:connected-moments}
  c_{k_1,\ldots,k_n}=\sum_{\cM\in \bM_{k_1,\ldots,k_n}^c} N^{V_{\circ}(\cM)-p+F(\cM)}.
\end{equation}
Thanks to the connected condition, this sum is a polynomial in $1/N$ as long as $n>1$. That is to say we have
\begin{equation}
    c_{k_1,\ldots,k_n}=\sum_{g\ge 0}N^{2-n-2g}c_{k_1,\ldots,k_n}^{[g]}.
\end{equation}
This last equation is shown starting from \eqref{eq:connected-moments} and again using $V_{\bullet}+V_{\circ}(\cM)-k+F(\cM)=2-2g(\cM)$ with $V_{\bullet}=n$.\\

\noindent{\bf Large $N$ limit of moments of a Wishart matrix.} Using \eqref{eq:genus-exp-moments}, one can study the large $N$ limit of the moments of order $k$ of a Wishart matrix, that is one can compute the limit
\begin{equation}
   \lim_{N\rightarrow \infty} \frac1N m_k=m^{[0]}_k.
\end{equation}
This limit is given by the number of planar, labeled, bicolored combinatorial maps with one black vertex and $k$ edges. The number of such maps is given by the Catalan number\footnote{Note that one obtains Catalan numbers when the ratio parameter is set to $c=1$, however for general values of $c$ one obtains the Narayana statistics on trees, that is polynomials in $c$ whose coefficients are Narayana numbers \cite{DR-NarayanaWishart}.} $C_k$ so that $m^{[0]}_k=C_k=\frac1{k+1}\binom{2k}{k}$. This allows to compute the large $N$ limit $W_{0,1}(x)$ of the moment generating function of the Wishart matrix
\begin{equation}\label{eq:W01-Wishart-expression}
  W_{0,1}(x):=\lim_{N\rightarrow \infty}\frac1N \bE\left( \Tr\left((x-W)^{-1} \right)\right)=\sum_{p\ge 0}\frac{m^{[0]}_p}{x^{p+1}} =\frac{x-\sqrt{x^2-4x}}{2x}.
\end{equation}
This last quantity is the Stieltjes transform of the limiting eigenvalues density of the Wishart matrix.
The knowledge of $W_{0,1}(x)$ allows in principle\footnote{In this specific case one can recover explicitly the limiting eigenvalue density via the inverse transformation. However in general it can be more tedious to compute the inverse transform. In the cases where the equation determining $W_{0,1}$ is an algebraic equation, one can deduce a system of polynomial equations on two quantities $u(x), v(x)$, one of them being (proportional to) the large $N$ limit of the eigenvalue density $\rho_{0,1}(x)$. We illustrate this fact in the later Remarks \ref{rem:poly-density}, \ref{rem:poly-density2}.} to recover the limiting eigenvalues density via the inverse transformation.   \\

\noindent{\bf Schwinger-Dyson equation method.} In this part we use an alternative method to compute $W_{0,1}(x)$. We use the Wishart case as a pedagogical example. The Schwinger-Dyson equation method relies on the use of the simple identity
\begin{equation}
    \sum_{a,b=1}^N\int \frac{N^{N^2}}{(2i\pi)^{N^2}}\mathrm{d}X^{\dagger}\mathrm{d}X \partial_{X^{\dagger}_{ab}}\left((X^{\dagger}S_1^{k})_{ab}e^{-N\Tr(XX^{\dagger})}\right)=0,
\end{equation}
after computing the derivatives explicitly we obtain the following set of relations between moments
\begin{equation}\label{eq:SD-}
    \sum_{\substack{p_1,p_2\ge 0 \\ p_1+p_2=k}}m_{p_1,p_2}-Nm_{k+1}=0.
\end{equation}
In order to continue this computation we define the $n$-points resolvents $\overline{W}_n(x_1,\ldots,x_n)$ and their connected counterpart $W_n(x_1,\ldots,x_n)$
\begin{align}
  \overline{W}_n(x_1,\ldots,x_n)&:=\bE\left(\prod_{i=1}^n\Tr\left((x_i-S_1)^{-1}\right)\right)=\sum_{p_1,\ldots,p_n\ge0}\frac{m_{p_1,\ldots,p_n}}{x_1^{p_1+1}\ldots x_n^{p_n+1}}  \\
  W_n(x_1,\ldots,x_n)&=\sum_{p_1,\ldots,p_n\ge0}\frac{c_{p_1,\ldots,p_n}}{x_1^{p_1+1}\ldots x_n^{p_n+1}}.
\end{align}
Note that we will often name the $n$-points resolvents and their connected counterpart simply resolvents, unless the context makes it unclear which object we are discussing. $W_{0,1}(x)$ is (up to normalization) the large $N$ limit of $W_1(x)$. We have the relation 
\begin{equation}\label{eq:diconnect-to-connect}
    \overline{W}_n(x_1,\ldots,x_n)=\sum_{K\vdash\{1,\ldots,n\}}\prod_{K_i\in K}W_{\mid K_i\mid}(x_{K_i}),
\end{equation}
where we used the notation $x_{K_i}=\{x_j\}_{j\in K_i}$. The above relation is inherited from the moment-cumulant relation of equation \eqref{eq:mom-cum-for-traces}.
\vspace{2mm}
\begin{remark}
Note that $\overline{W}_1(x)=W_1(x)$.
\end{remark}
\vspace{2mm}
With these definitions in mind, one considers the equality
\begin{equation}
    \sum_{k\ge 0}\frac1{x^{k+1}}\left(\sum_{\substack{p_1,p_2\ge 0 \\ p_1+p_2=k}}m_{p_1,p_2}-Nm_{k+1}\right)=0,
\end{equation}
leading after some rewriting to
\begin{equation}\label{eq:Wishart-1pt-res-disconnect}
    \overline{W}_2(x,x)-NW_1(x)+N^2/x=0,
\end{equation}
or only in terms of the connected resolvents
\begin{equation}\label{eq:Wishart-1pt-res}
    W_1(x)^2+W_2(x,x)-NW_1(x)+N^2/x=0.
\end{equation}
The (connected) resolvents inherit a $1/N$ expansion from the expansion of the cumulants,
\begin{equation}\label{eq:exp-conn-resolvents}
  W_n(x_1,x_2,\ldots,x_n)=\sum_{g\ge0}N^{2-2g-n}W_{g,n}(x_1,x_2,\ldots,x_n)
\end{equation}
and thus we have 
\begin{equation}
    W_1(x)=\sum_{g\ge 0} N^{1-2g}W_{g,1}(x),\quad 
    W_2(x,x)=\sum_{g\ge 0}N^{-2g}W_{g,2}(x,x).
\end{equation}
In the large $N$ limit equation \eqref{eq:Wishart-1pt-res} reduces to an equation on $W_{0,1}(x)$,
\begin{equation}\label{eq:large-N-Wishart-1pt-res}
    xW_{0,1}(x)^2-xW_{0,1}(x)+1=0.
\end{equation}
From which we select the solution which is analytic at infinity thus recovering expression \eqref{eq:W01-Wishart-expression}.\\

\begin{remark}\label{rem:poly-density}
 From this last equation we can obtain a polynomial equation on $\rho_{0,1}(x)$, that is the corresponding limiting eigenvalue density. To this aim, one introduces the two following operators acting on functions,
 \begin{align}
    &\delta f(x)=\lim_{\epsilon \rightarrow 0^+}f(x+i\epsilon)-f(x-i\epsilon)\\
    &\s f(x)=\lim_{\epsilon \rightarrow 0^+}f(x+i\epsilon)+f(x-i\epsilon).
 \end{align}
 We have the following \emph{polarization} property, that is for two functions $f_1, f_2$, we have
 \begin{align}
     &\delta(f_1f_2)(x)=\frac12(\delta f_1(x)\s f_2(x)+\s f_1(x)\delta f_2(x))\\
     &\s (f_1f_2)(x)=\frac12(\delta f_1(x)\delta f_2(x)+\s f_1(x)\s f_2(x))
 \end{align}
 Starting from equation \eqref{eq:large-N-Wishart-1pt-res} one deduces the two equalities
 \begin{align}
    &\delta(xW_{0,1}(x)^2-xW_{0,1}(x)+1)=0\\
    &\s (xW_{0,1}(x)^2-xW_{0,1}(x)+1)=0.
 \end{align}
 After using the polarization formula, these equations boil down to the system on $u(x):=\s W_{0,1}(x)$ and $v(x):=\delta W_{0,1}(x)$
 \begin{align}
    &xu(x)-x=0\\
    &\frac{x}{2}(u(x)^2+v(x)^2)-xu(x)+2=0.
 \end{align}
 This in turn leads to $\rho_{0,1}(x)=\frac1{2i\pi}v(x)=\frac1{2\pi}\sqrt{\frac{x-4}{x}}$, where we choose the solution $v(x)$ that leads to a positive and normalized density.
\end{remark}

\section{Loop equations for the product of two Ginibre matrices}\label{sec:loop-eq-prod}
In this section we consider the problem of computing $W_{0,1}(x)$, $W_{0,2}(x_1,x_2)$ and $W_{1,1}(x)$ for a matrix $S_2=X_1X_1^{\dagger}X_2^{\dagger}X_2$ with $X_1, X_2$ two random $N\times N$ complex matrices with normal entries of mean zero. We compute these quantities by exclusive use of Schwinger-Dyson equation techniques. More generally, we obtain the general equations satisfied by any $W_{g,n}$ for $(g,n)\ge (0,1)$.

In the first subsection, we briefly explain the combinatorics underlying the computation of the moments of the matrix $S_2$ that justifies the existence of a $1/N$ expansion for the $W_{g,n}$.
In the second subsection we study in details the corresponding Schwinger-Dyson equations and obtain the loop equations satisfied by $W_{0,1}(x)$, $W_{0,2}(x_1,x_2)$ and $W_{1,1}(x)$ in this context. We show in particular that the loop equation satisfied by $W_{0,1}(x)$ is an algebraic equation of degree $3$ in $W_{0,1}$. Finally we describe the loop equations satisfied by any $W_{g,n}$.
\vspace{2mm}
\subsection{Combinatorics of the moments of $S_2$ and existence of $1/N$ expansion}
\begin{figure}
    \centering
    \includegraphics[scale=0.8]{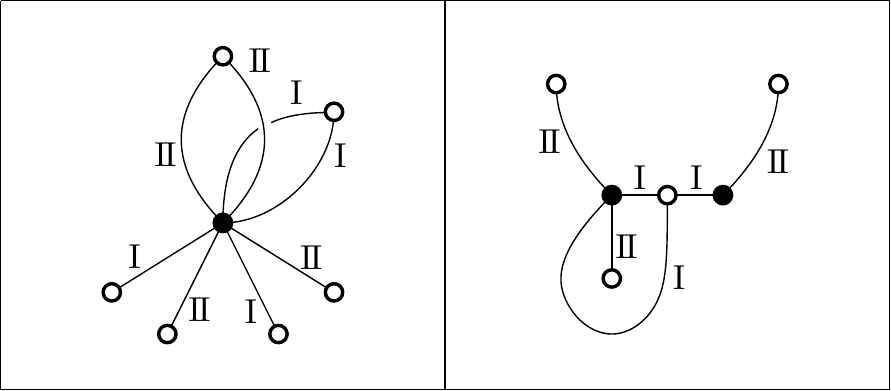}
    \caption{Left: Example of a map with two types of edge contributing to the computation of $m_4$. Right: Example of a map with two types of edge contributing to the computation of $m_{2,1}$ and $c_{2,1}$.}
    \label{fig:colored-maps-examples}
\end{figure}

We describe here the combinatorics of the moments of the matrix $S_2$. This is a crucial point as this underlying combinatorics allows us to show that the cumulants of the random variables $\left\{\Tr(S_2^{i})\right\}_{i=0}^{\infty}$ have a $1/N$ expansion.
In the subsequent developments, we keep the same notation for the moments $m_k$, $m_{k_1,\ldots, k_n}$  but it should be clear that in this section and the following, the moments we consider are the moments of the matrix $S_2$, and that is so, in both the one trace case, and the multiple traces case. We have
\begin{equation}
    m_k=\bE\left(\Tr(S_2^k)\right), \quad m_{k_1,\ldots,k_n}=\bE\left(\prod_{i=1}^n\Tr(S_2^{k_i})\right),
\end{equation}
where the expectation is taken with respect to the density
\begin{equation}\label{eq:two-mat-density}
 \mathrm{d}\mu(X_1,X_2)=\left(\frac{N^{N^2}}{(2i\pi)^{N^2}}\right)^2e^{-N\Tr(X_1X_1^{\dagger})}e^{-N\Tr(X_2X_2^{\dagger})}\mathrm{d}X_1^{\dagger}\mathrm{d}X_1\mathrm{d}X_2^{\dagger}\mathrm{d}X_2.
\end{equation}
By using the Wick-Isserlis theorem, it is possible to give a combinatorial interpretation to the moments of $S_2$ (see for instance \cite{DLN}). The moments $m_k$ of $S_2$ write as a sum over combinatorial maps with one black vertex, $2k$ edges of two different types, type \RNum{1} and type \RNum{2}, such that there are $k$ edges of type \RNum{1} and $k$ edges of type \RNum{2}. Moreover the type of the edge alternates when going around the black vertex. Finally the white vertices can only be incident to edges of one given type. See Fig. \ref{fig:colored-maps-examples} for examples. 

We denote the set  made of these maps by $\bM_{2k}(2)$. In terms of permutations, these maps are such that $\sigma_{\bullet}=(12\ldots2k)$ and the action of $\sigma_{\circ}$ on the set of edges $E=\{1,2,3,4,\ldots,2k\}$ factorizes over the odd and even subsets $E_o=\{1,3,5,\ldots, 2k-1\}, E_e=\{2,4,6,\ldots,2k\}$. More formally we have the  decomposition
\begin{equation}
    \sum_{\cM\in\bM_{2k}(2)}N^{V_{\circ}(\cM)-2k+ F(\cM)}.
\end{equation}
Similarly, for moments of order $k_1,\ldots,k_n$, we have the set of maps $\bM_{2k_1, 2k_2, \ldots, 2k_n}(2)$, such that there are $n$ black vertices with degree distribution $2k_1, 2k_2, \ldots, 2k_n$ and a total of $p=2\sum_i k_i$ edges. Types of edge alternate around each black vertex, and white vertices can only be incident to edges of the same type see Fig. \ref{fig:colored-maps-examples} for examples. We then have the decomposition
\begin{equation}
    m_{k_1,\ldots, k_n}=\sum_{\cM \in\bM_{2k_1, 2k_2, \ldots, 2k_n}(2)} N^{V_{\circ}(\cM)-p+F(\cM)}.
\end{equation}
Similarly we can express the cumulants $c_{k_1,\ldots, k_n}$ for the family of random variables $\left\{\Tr(S_2^{i})\right\}_{i=0}^{\infty}$ as a sum over the set of connected maps $\bM^c_{2k_1, 2k_2, \ldots, 2k_n}(2)$ 
\begin{equation}
    c_{k_1,\ldots, k_n}=\sum_{\cM \in\bM^c_{2k_1, 2k_2, \ldots, 2k_n}(2)} N^{V_{\circ}(\cM)-p+F(\cM)}.
\end{equation}
The connected condition ensures that the $c_{k_1,\ldots, k_n}$ have a $1/N$ expansion for $n\ge1$. This $1/N$ expansion as well as the definition of $c_{k_1,\ldots, k_n}$ as the cumulants of the family $\left\{\Tr(S_2^{i})\right\}_{i=0}^{\infty}$ ensure that the resolvents for the matrix $S_2$ have the same structural properties than the resolvents of the Wishart matrix in equations \eqref{eq:diconnect-to-connect}, \eqref{eq:exp-conn-resolvents}, that is we also have for the matrix $S_2$
\begin{align}
    \label{eq:res-con-to-disc-S_2}&\overline{W}_n(x_1,\ldots,x_n)=\sum_{K\vdash\{1,\ldots,n\}}\prod_{K_i\in K}W_{\mid K_i\mid}(x_{K_i}),\\
    \label{eq:res-exp-S_2}&W_n(x_1,x_2,\ldots,x_n)=\sum_{g\ge0}N^{2-2g-n}W_{g,n}(x_1,x_2,\ldots,x_n).
\end{align}
\subsection{Equation on $W_{1}$ and $W_{0,1}$}
We now want to write Schwinger-Dyson equations for the moments of the matrix $S_2$ in order to obtain the loop equations for the resolvents. We start with the set of identities
\begin{align}\label{eq:FirstorderSDE}
&\sum_{a,b=1}^N\int \mathrm{d}X_1\mathrm{d}X_1^{\dagger} \mathrm{d}X_2\mathrm{d}X_2^{\dagger}\frac{\partial}{\partial X_{1,ab}^{\dagger}}\left(\bigl[X_1^{\dagger}X_2^{\dagger}X_2 S_2^{k}\bigr]_{ab} e^{-N\Tr(X_1X_1^{\dagger})}e^{-N\Tr(X_2X_2^{\dagger})}\right) =0 \displaybreak[1] \\
&\sum_{a,b=1}^N\int \mathrm{d}X_1\mathrm{d}X_1^{\dagger} \mathrm{d}X_2\mathrm{d}X_2^{\dagger}\frac{\partial}{\partial X_{2,ab}^{\dagger}}\left(\bigl[S_2^{k}X_1X_1^{\dagger}X_2^{\dagger}\bigr]_{ab} e^{-N\Tr(X_1X_1^{\dagger})}e^{-N\Tr(X_2X_2^{\dagger})}\right) =0.
\end{align}
After evaluating explicitly the action of the derivatives, we obtain relations,
\begin{align}
&\sum_{\substack{p_1+p_2=k \\ p_1,p_2\ge 0}} \bE\left( \Tr(S_2^{p_1})\Tr(S_2^{p_2}X_2^{\dagger}X_2)\right)-N\bE\left( \Tr(S_2^{k+1})\right)=0\\
&\sum_{\substack{p_1+p_2=k \\ p_1,p_2\ge 0}} \bE\left( \Tr(S_2^{p_1}X_1X_1^{\dagger})\Tr(S_2^{p_2})\right)-N\bE\left( \Tr(S_2^{k+1})\right)=0,
\end{align}
where for both equation, the first term comes from the evaluation of the derivative on the monomial, while the second term comes from the evaluation of the derivative on the exponential factor. Note however that these equations contain mixed terms of the form $\bE\left( \Tr(S_2^{p_1})\Tr(S_2^{p_2}X_2^*X_2)\right)$ and $\bE\left( \Tr(S_2^{p_1}X_1X_1^*)\Tr(S_2^{p_2})\right)$ that cannot be expressed in terms of the moments of $S_2$. Thus these two equations do not close on the set of moments of $S_2$. In order to obtain a set of relations that closes over the set of moments of $S_2$, we consider another identity involving higher derivatives. This is,
\begin{align}
\int  \mathrm{d}X_1\mathrm{d}X_1^{\dagger} \mathrm{d}X_2\mathrm{d}X_2^{\dagger}\frac{\partial}{\partial X_{1,ab}^{\dagger}}\frac{\partial}{\partial X_{2,bc}^{\dagger}}\left( \bigl[X_1^{\dagger}X_2^{\dagger}X_2 S_2^{k}X_1X_1^{\dagger}X_2^{\dagger}\bigr]_{ac}e^{-N\Tr(X_1X_1^{\dagger})}e^{-N\Tr(X_2X_2^{\dagger})}\right)=0,
\end{align}
where we sum over repeated indices. After some additional algebra to evaluate the action of both derivative operators, one gets relations between moments and additional mixed quantities
\begin{multline}\label{eq:SecondorderSDE}
 \sum_{\substack{p_1+p_2+p_3 =k+1 \\ p_1,p_2,p_3 \ge 0}} \bE\bigl(\Tr(S_2^{p_1})\Tr(S_2^{p_2}) \Tr(S_2^{p_2})\bigr)+\frac{(k+1)(k+2)}{2}\bE\bigl(\Tr(S_2^{k+1})\bigr) \\
 -N\sum_{\substack{p_1+p_2=k+1 \\ p_1,p_2\ge 0}} \Bigl[ \bE\bigl( \Tr(S_2^{p_1})\Tr(S_2^{p_2}X_2^*X_2)\bigr) +\bE\bigl( \Tr(S_2^{p_1}X_1X_1^*)\Tr(S_2^{p_2})\bigr) \Bigr] \\
 +N^2 \bE\left( \Tr(S_2^{k+2}) \right)=0,
\end{multline}
where the first and second terms are obtained from the action of both derivatives operators on the monomial $\bigl[X_1^{\dagger}X_2^{\dagger}X_2 S_2^{k}X_1X_1^{\dagger}X_2^{\dagger}\bigr]_{ac}$. The third term that involves mixed quantities is obtained by acting with one derivative operator on the monomial, while acting with the other derivative operator on the exponential factor. The last term is obtained from the action of both derivative operator on the exponential factor. These equations contain the mixed quantities already present in \eqref{eq:FirstorderSDE}. Thus we can use \eqref{eq:FirstorderSDE} to get rid of these terms in \eqref{eq:SecondorderSDE}. This leads to the  equations on moments
\begin{align}
\sum_{\substack{p_1+p_2+p_3 =k+1\\ p_1,p_2,p_3 \ge 0}} \bE\left( \Tr(S_2^{p_1})\Tr(S_2^{p_2}) \Tr(S_2^{p_2})\right) +\frac{(k+1)(k+2)}{2}\bE\left(\Tr(S_2^{k+1})\right)-N^2 \bE\left(\Tr(S_2^{k+2}) \right)=0,
\end{align}
which is trilinear in the traces of $S_2$. Notice that this family of equations extends to the value ``$k=-1$" by replacing the monomial $\bigl[X_1^{\dagger}X_2^{\dagger}X_2 S_2^{k}X_1X_1^{\dagger}X_2^{\dagger}\bigr]_{ac}$ by $\bigl[X_1^{\dagger}X_2^{\dagger}\bigr]_{ac}$. Therefore we allow ourselves to set $k=k-1$ and to use our moments notation to get
\begin{align}
\sum_{\substack{p_1+p_2+p_3 =k\\ p_1,p_2,p_3 \ge 0}} m_{p_1,p_2,p_3}+\frac{k(k+1)}{2}m_k -N^2 m_{k+1}=0.
\end{align}
We then multiply the above equation by $\frac1{x^{k+1}}$ and sum over $k\ge 0$ in order to get an equation on the resolvents
\begin{equation}
\sum_{k\ge 0}\sum_{\substack{p_1+p_2+p_3 =k\\ p_1,p_2,p_3 \ge 0}}\frac{m_{p_1,p_2,p_3}}{x^{k+1}}+\sum_{k\ge 0}\frac{k(k+1)}{2}\frac{m_k}{x^{k+1}} -N^2 \frac{m_{k+1}}{x^{k+1}}=0,
\end{equation}
which after a few manipulations rewrites
\begin{equation}\label{eq:1pt-equation}
x^2\overline W_3(x,x,x)+x\partial_xW_1(x)+\frac12x^2\partial_x^2W_1(x)-N^2xW_1(x)+N^3 =0.
\end{equation}
Note the interesting structural replacement of $\overline W_2(x,x)$ appearing in \eqref{eq:Wishart-1pt-res-disconnect} by $\overline W_3(x,x,x)$ and the appearance of a derivative term.
Then we know from \eqref{eq:res-con-to-disc-S_2}, \eqref{eq:res-exp-S_2} that  $\overline W_3(x,x,x)= N^3 W_{0,1}(x)^3+O(N)$ and $W_1(x)= N W_{0,1}(x)+O(1/N)$. Therefore we obtain the equation on $W_{0,1}(x)$
\begin{equation}\label{eq:W01-equation}
x^2W_{0,1}(x)^3-xW_{0,1}(x)+1=0.
\end{equation}
This last equation relates to the equation satisfied by the generating function $G(u)$ of particular Fuss-Catalan numbers \cite{Fuss-original, Mlotkowski2010, Rivass-FussCatalan}, $uG(u)^3-G(u)+1=0$ through the change of variables $W_{0,1}(x)=\frac1{x}G(1/x)$. Consequently we have
\begin{equation}
    W_{0,1}(x)=\sum_{p\ge0} \frac{C_p[3]}{x^{p+1}},
\end{equation}
where $C_p[D]$ are the Fuss-Catalan numbers of order $D$, the usual Catalan numbers $C_p$ being the Fuss-Catalan numbers of order $2$, that is $C_p=C_p[2]$, and have the binomial coefficient form
\begin{equation}
    C_p[D]=\frac1{(D-1)p+1}\binom{Dp}{p}.
\end{equation}

An explicit form of $W_{0,1}(x)$ can be written as follows. First define 
\begin{equation}
    K_{\pm}(u)=(\sqrt{1+u}\pm\sqrt{u})^{1/3},
\end{equation}
then $G(u)$ writes
\begin{equation}
    G(u)=\frac{K_{+}\left(-\frac{27u}{4}\right)-K_{-}\left(-\frac{27u}{4}\right)}{\sqrt{-3u}}.
\end{equation}
Finally one has
\begin{equation}
    W_{0,1}(x)=\frac1{x}G\left(\frac1{x}\right).
\end{equation}
We study the solutions and the structure of \eqref{eq:W01-equation} from a geometric perspective in the next sections.

\begin{figure}
    \centering
    \includegraphics[scale=0.52]{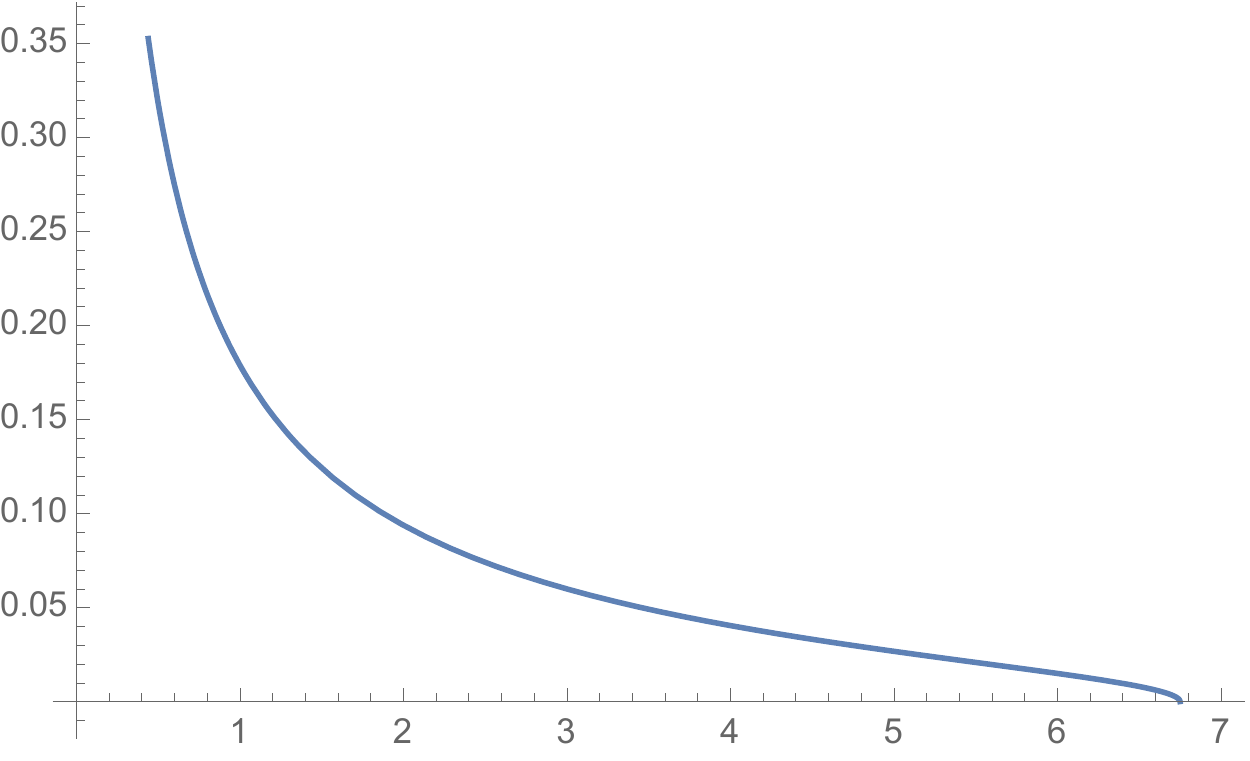}
    \caption{Plot of the eigenvalue density of the matrix $S_2$ in the large $N$ regime.}
    \label{fig:eigen-density-product}
\end{figure}

\begin{remark}\label{rem:poly-density2}
Though in principle we need to first focus on the cut structure of $W_{0,1}$ to use the arguments that follow, we will in this remark content ourselves with a formal computation. Starting from equation \eqref{eq:W01-equation} we can also obtain a polynomial equation satisfied by the corresponding density by using the $\delta, \s$ operators along the cut. Indeed with a similar method to that in Remark \ref{rem:poly-density} we have the equalities
\begin{align}
    &\delta(x^2W_{0,1}(x)^3-xW_{0,1}(x)+1)=0\\
    &\s(x^2W_{0,1}(x)^3-xW_{0,1}(x)+1)=0.
\end{align}
This leads, using the same previously used notations, to the system
\begin{align}
    &\frac{x^2}{4}(3u(x)^2+v(x)^2)-x=0\\
    &\frac{x^2}{4}(u(x)^3+3v(x)^2u(x))-xu(x)+2=0
\end{align}
which can be solved and leads to 
\begin{equation}\label{63}
    \rho_{0,1}(x)=\frac1{2i\pi}v(x)=\frac1{2 \pi }\sqrt{\frac{\left(\sqrt{81-12 x}+9\right)^{2/3}}{2^{2/3} \sqrt[3]{3} x^{4/3}}+\frac{2^{2/3}
   \sqrt[3]{3}}{\left(\left(\sqrt{81-12 x}+9\right) x\right)^{2/3}}-\frac{2}{x}},
\end{equation}
which is supported on $(0,27/4]$, see the plot of the distribution on Fig. \ref{fig:eigen-density-product}. Notice that this result can also be obtained by computing the free multiplicative product of two Mar\u cenko-Pastur distribution of parameters $c_{1,2}=1$. A functional form equivalent to (\ref{63}) is given in \cite{penson2011product}.
\end{remark}
Equation \eqref{eq:1pt-equation} possesses a $\frac1{N}$ expansion. This expansion results in a set of relations between $W_{g,1}(x)$, $W_{g',2}(x,x)$ and $W_{g'',3}(x,x,x)$. Indeed we have 
\begin{multline}
0=x^2\Bigl[\frac1N\sum_{g\ge 0}N^{-2g}W_{g,3}(x,x,x)+3N\sum_{g_1,g_2\ge 0}N^{-2(g_1+g_2)}W_{g_1,1}(x)W_{g_2,2}(x,x) \\
+N^3\sum_{g_1,g_2,g_3\ge 0}N^{-2(g_1+g_2+g_3)}W_{g_1,1}(x)W_{g_2,1}(x)W_{g_3,1}(x)\Bigr]\\
+xN\sum_{g\ge 0} N^{-2g}\partial_x W_{1,g}(x)+\frac{N}{2}x^2\sum_{g\ge 0}N^{-2g}\partial_x^2 W_1(x)-N^3x\sum_{g\ge 0}N^{-2g}W_{g,1}(x)+N^3.
\end{multline}
By collecting the coefficient of $N^{3-2g}$, we obtain the following tower of equations
\begin{multline}\label{eq:order-g-1pt}
0=x^2\left(W_{g-2,3}(x,x,x)+3\sum_{g_1+g_2=g-1}W_{g_1,1}(x)W_{g_2,2}(x,x)+\sum_{g_1+g_2+g_3=g}W_{g_1,1}(x)W_{g_2,1}(x)W_{g_3,1}(x)\right)\\
+x\partial_xW_{g-1,1}(x)+\frac{x^2}{2} \partial_x^2W_{g-1,1}(x)-xW_{g,1}(x)+P_{g,1}(x),
\end{multline}
where we have $P_{g,1}(x)=\delta_{g,0}$. In particular, the coefficient of $N^3$ of equation \eqref{eq:order-g-1pt} produces equation \eqref{eq:W01-equation}.
The coefficient of $N$ produces an equation on the next-to-leading order $W_{1,1}(x)$ also involving $W_{0,1}(x)$ and $W_{0,2}(x,x)$
\begin{equation}\label{eq:1pt-NLO}
3x^2W_{0,1}(x)W_{0,2}(x,x)+3x^2W_{0,1}(x)^2W_{1,1}(x)+x\partial_xW_{0,1}(x)+\frac{x^2}{2} \partial_x^2W_{0,1}(x)-xW_{1,1}(x)=0.
\end{equation}
More generally, the coefficient of $N^{3-2g}$ for a fixed value of $g$ produces the equation for $W_{g,1}(x)$ in terms of the functions $W_{g',n'}$ such that $2-2g-1<2-2g'-n'$ and $n'\le 3$.  \\
\subsection{Equation for $W_2(x_1,x_2)$}
In this section we use Schwinger-Dyson equation techniques to obtain a loop equation for $W_2(x_1,x_2)$. We start with slightly different identities that involve an additional trace insertion $\Tr(S_2^q)$. This allows us to access relations between more general moments.\\

\noindent{\bf Schwinger-Dyson equations and loop equation for $W_2(x_1,x_2)$ and $W_{0,2}(x_1,x_2)$.} Consider the vanishing integrals of total derivatives
\begin{align}
\label{eq:2pttotalder1}&\int \mathrm{d}X_1\mathrm{d}X_1^{\dagger} \mathrm{d}X_2\mathrm{d}X_2^{\dagger}\frac{\partial}{\partial X_{1,ab}^{\dagger}}\left(\bigl[X_1^{\dagger}X_2^{\dagger}X_2 S_2^{k+1}\bigr]_{ab} \Tr(S_2^q)e^{-N\Tr(X_1X_1^{\dagger})}e^{-N\Tr(X_2X_2^{\dagger})}\right) =0 \\
\label{eq:2pttotalder2}&\int \mathrm{d}X_1\mathrm{d}X_1^{\dagger} \mathrm{d}X_2\mathrm{d}X_2^{\dagger}\frac{\partial}{\partial X_{2,ab}^{\dagger}}\left(\bigl[S_2^{k+1}X_1X_1^{\dagger}X_2^{\dagger}\bigr]_{ab} \Tr(S_2^q)e^{-N\Tr(X_1X_1^{\dagger})}e^{-N\Tr(X_2X_2^{\dagger})}\right) =0,
\end{align}
and the higher derivative one
\begin{align}
\label{eq:2pttotalder-higher}\int \mathrm{d}X_1\mathrm{d}X_1^{\dagger} \mathrm{d}X_2\mathrm{d}X_2^{\dagger}\frac{\partial}{\partial X_{1,ab}^{\dagger}}\frac{\partial}{\partial X_{2,bc}^{\dagger}}\left( \bigl[X_1^{\dagger}X_2^{\dagger}X_2 S_2^{k}X_1X_1^{\dagger}X_2^{\dagger}\bigr]_{ac}\Tr(S_2^q)e^{-N\Tr(X_1X_1^{\dagger})}e^{-N\Tr(X_2X_2^{\dagger})}\right)=0,
\end{align}
where repeated indices are summed. After evaluating explicitly the derivatives, the two first equations \eqref{eq:2pttotalder1} and \eqref{eq:2pttotalder2} lead to 
\begin{align}
\label{eq:2ptSDE1}&\sum_{\substack{p_1+p_2=k+1 \\ \{p_i\ge 0\}}}\bE\left(\Tr(S_2^{p_1})\Tr(S_2^{p_2}X_2^{\dagger}X_2)\Tr(S_2^q) \right)+q\bE\left( \Tr(S_2^{k+q+1}X_2^{\dagger}X_2)\right)-N\bE\left( \Tr(S_2^{k+2})\Tr(S_2^q)\right)=0 \\
\label{eq:2ptSDE2}&\sum_{\substack{p_1+p_2=k+1 \\ \{p_i\ge 0\}}}\bE\left(\Tr(S_2^{p_1}X_1X_1^{\dagger}) \Tr(S_2^{p_2})\Tr(S_2^q)\right)+q\bE\left(\Tr(S_2^{k+q+1}X_1X_1^{\dagger}) \right)-N\bE\left(\Tr(S_2^{k+2})\Tr(S_2^q) \right)=0,
\end{align}
where the first term of both equations \eqref{eq:2pttotalder1} and \eqref{eq:2pttotalder2} is obtained from the action of the derivative operator on the non-traced monomial. The second term is obtained \textit{via} the action of the derivative operator on the traced monomial term $\Tr(S_2^q)$. The third term comes from the action of the derivative operator on the exponential factor. These two equations involve mixed terms and cannot be written solely in terms of the moments of $S_2$.
Meanwhile, the higher derivative equation \eqref{eq:2pttotalder-higher} leads to
\begin{multline}\label{eq:2ptSDE-higher}
\sum_{\substack{p_1+p_2+p_3=k+1 \\ \{p_i\ge 0\}}}\bE\left(\Tr(S_2^{p_1})\Tr(S_2^{p_2})\Tr(S_2^{p_3})\Tr(S_2^q) \right)+\frac{(k+1)(k+2)}{2}\bE\left( \Tr(S_2^{k+1})\Tr(S_2^q) \right) \\
- N\sum_{\substack{p_1+p_2=k+1 \\ \{p_i\ge 0\}}}\left[ \bE\left(\Tr(S_2^{p_1})\Tr(S_2^{p_2}X_2^{\dagger}X_2)\Tr(S_2^q)\right) + \bE\left( \Tr(S_2^{p_1}X_1X_1^{\dagger})\Tr(S_2^{p_2})\Tr(S_2^q) \right) \right] \\
+N^2\bE\left(\Tr(S_2^{k+2})\Tr(S_2^q) \right)+2\sum_{\substack{p_1,p_2\ge 0\\ p_1+p_2=k+1}}q\bE\left(\Tr(S_2^{p_1})\Tr(S_2^{p_2+q}) \right)
+\sum_{n=1}^q q \bE\left( \Tr(S_2^{k+1+n})\Tr(S_2^{n})\right) \\
- Nq\left[ \bE\left( \Tr(S_2^{q+k+1}X_2^{\dagger}X_2) \right) + \bE\left( \Tr(S_2^{q+k+1}X_1X_1^{\dagger})\right) \right] =0,
\end{multline}
where the two first terms come from the action of both derivatives operators on the non-traced monomial. Each term of the second line comes from the action of one of the derivative on the exponential factor and of the other on the non-traced monomial. The first term of the third line of \eqref{eq:2ptSDE-higher} comes from the action of both derivatives on the exponential factor. The second term of the third line is obtained as a sum of the action of the $X_1^{\dagger}$ (resp. $X_2^{\dagger}$) derivative on the non-traced monomial and the action of the $X_2^{\dagger}$ (resp. $X_1^{\dagger}$) derivative on the traced monomial $\Tr(S_2^q)$. The last term of the third line is obtained from the action of both derivative  operators on the traced monomial. Finally the two terms of the fourth line of \eqref{eq:2ptSDE-higher} are obtained by the action of $\partial_{X_{1,ab}^{\dagger}}$ (resp. $\partial_{X_{2,bc}^{\dagger}}$) on the traced monomial and $\partial_{X_{2,bc}^{\dagger}}$ (resp. $\partial_{X_{1,ab}^{\dagger}}$) on the exponential factor.
Combining equations \eqref{eq:2ptSDE1}, \eqref{eq:2ptSDE2} and \eqref{eq:2ptSDE-higher}, rewriting some of the sums in a nicer way and using our moments notation we obtain
\begin{multline}\label{eq:2pt-moment-relation}
\sum_{\substack{p_1+p_2+p_3=k+1 \\ \{p_i\ge 0\}}}m_{p_1,p_2,p_3,q} +\frac{(k+1)(k+2)}{2}m_{k+1,q} - N^2 m_{k+2,q}
+\sum_{\substack{p_1,p_2\ge 0\\ p_1+p_2=k+1}}qm_{p_1,p_2+q}\\
+\sum_{\substack{p_1,p_2 \ge 0 \\ p_1+p_2= k+q+1}}q m_{p_1,p_2} =0.
\end{multline}
After performing the shift $k\rightarrow k-1$ in \eqref{eq:2pt-moment-relation}, we multiply \eqref{eq:2pt-moment-relation} by $\frac1{x_1^{k+1}x_2^{q+1}}$, and sum over $k, q\ge 0$. Doing so we obtain the equation 
\begin{align}
0=&\overline W_4(x_1,x_1,x_1,x_2)+\frac1{x_1}\partial_{x_1}\overline W_2(x_1,x_2)+\frac1{2}\partial_{x_1}^2\overline W_2(x_1,x_2) - \frac{N^2}{x_1}\overline W_2(x_1,x_2)-N^2A_2(x_1,x_2) \\
&+\frac{1}{x_1^2}\partial_{x_2}\left( x_1x_2\frac{\overline W_2(x_1,x_1)-\overline W_2(x_1,x_2)}{x_1-x_2} \right)+\frac1{x_1^2}\partial_{x_2}\left( \frac{x_1x_2\overline W_2(x_1,x_1)-x_2^2\overline W_2(x_2,x_2)}{x_1-x_2}\right).
\end{align}
with $A_2(x_1,x_2)=-\frac{N}{x_1^2}W_1(x_2)$. We re-express this equation in terms of the connected resolvents to obtain
\begin{multline}\label{eq:connected-2pt-resolvent-equation}
W_4(x_1,x_1,x_1,x_2)+3W_1(x_1)W_3(x_1,x_1,x_2)+3W_2(x_1,x_2)W_2(x_1,x_1)+3W_1(x_1)W_1(x_1)W_2(x_1,x_2)  \\
+\frac1{x_1}\partial_{x_1} W_2(x_1,x_2)+\frac1{2}\partial_{x_1}^2 W_2(x_1,x_2) - \frac{N^2}{x_1} W_2(x_1,x_2)+\frac1{x_1^2}\partial_{x_2}\left(x_1x_2\frac{W_2(x_1,x_1)-W_2(x_1,x_2)}{x_1-x_2}\right)\\
+\frac1{x_1^2}\partial_{x_2}\left(\frac{x_1x_2W_2(x_1,x_1)-x_2^2W_2(x_2,x_2)}{x_1-x_2}\right)+\frac1{x_1^2}\partial_{x_2}\left(x_1x_2\frac{W_1(x_1)W_1(x_1)-W_1(x_1)W_1(x_2)}{x_1-x_2}\right)  \\
+\frac1{x_1^2}\partial_{x_2}\left(\frac{x_1x_2W_1(x_1)W_1(x_1)-x_2^2W_1(x_2)W_1(x_2)}{x_1-x_2}\right)=0,
\end{multline}
where we used the fact that the terms factoring in front of $W_1(x_2)$ form the first loop equation \eqref{eq:1pt-equation}.
From this equation we can get an equation on $W_{0,2}$ by inserting the $1/N$ expansion of the resolvents appearing in \eqref{eq:connected-2pt-resolvent-equation} and collecting the coefficients of $N^2$. This equation involves only already computed quantities and can be re-expressed as
\begin{align}\label{eq:W02-equation}
\frac1{x_1}\left(3x_1W_{0,1}(x_1)^2-1 \right) W_{0,2}(x_1,x_2)+\frac1{x_1^2}\partial_{x_2}\left(x_1 x_2\frac{W_{0,1}(x_1)W_{0,1}(x_1)-W_{0,1}(x_1)W_{0,1}(x_2)}{x_1-x_2} \right) \nonumber \\
+\frac1{x_1^2}\partial_{x_2}\left( \frac{x_1x_2W_{0,1}(x_1)W_{0,1}(x_1)-x_2^2W_{0,1}(x_2)W_{0,1}(x_2)}{x_1-x_2}\right)=0.
\end{align}

\noindent{\bf First few relations between $c^{[0]}_{k}, c^{[0]}_{k_1,k_2}$.} One can extract relations between the $c^{[0]}_{k}, c^{[0]}_{k_1,k_2}$ from equation \eqref{eq:W02-equation}. These relations are obtained by expanding the equation at $x_1, x_2 = \infty$. The first few examples are
\begin{align}
    &3 c^{[0]}_0 c^{[0]}_1-c^{[0]}_{1,1}=0,\\
    &2 (c^{[0]}_1)^2+6 c^{[0]}_0 c^{[0]}_2-c^{[0]}_{1,2}=0,\\
    &6 c^{[0]}_1 c^{[0]}_2+9 c^{[0]}_0 c^{[0]}_3-c^{[0]}_{1,3}=0.
\end{align}
These relations allow to obtain the $c^{[0]}_{k_1,k_2}$ recursively knowing that $c^{[0]}_0,\  c^{[0]}_1=1$. We can check these first few relations combinatorially. For illustrative purposes we display the combinatorial maps interpretation of $3 c^{[0]}_0 c^{[0]}_1-c^{[0]}_{1,1}=0$
\begin{equation}
    3\, \left(\raisebox{-3mm}{\includegraphics[scale=0.65]{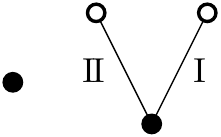}}\right)\quad - \quad \left(\raisebox{-4mm}{\includegraphics[scale=0.65]{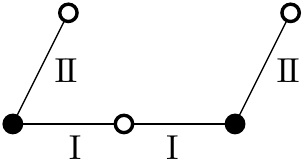}}\quad + \quad \raisebox{-3mm}{\includegraphics[scale=0.65]{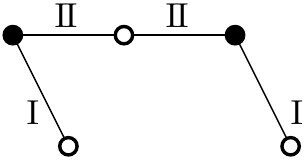}} \quad + \quad \raisebox{-4mm}{\includegraphics[scale=0.65]{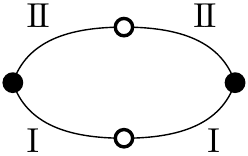}} \quad \right)\, =0.
\end{equation}
More generally, one has
\begin{equation}\label{eq:planar-bicumulant-equation}
    0=3\sum_{p_1+p_2+p_3=k-3}c^{[0]}_{p_1} c^{[0]}_{p_2}c^{[0]}_{p_3+1,q} - c^{[0]}_{k-1,q}+\sum_{m=0}^{k+q-2}q\ c^{[0]}_{k+q-m-2}c^{[0]}_{m}+\sum_{m=0}^{k-2}q\ c^{[0]}_{k-m-2}c^{[0]}_{m+q}.
\end{equation}\\
\subsection{General loop equations}
In this section we describe the general loop equations for $W_n(x_1,\ldots, x_n)$. Because of the use of higher derivatives for Schwinger-Dyson equations, the case of $W_3(x_1,x_2,x_3)$ is still special compared to the cases $W_{n<3}$. We thus give the corresponding Schwinger-Dyson equations in details before stating the corresponding loop equations. For the $W_{n>3}$ cases, the situation is very similar to the $W_{3}$ case. Therefore we refrain from presenting the detailed derivation, and only state the corresponding loop equations.\\

\noindent{\bf Loop and Schwinger-Dyson equations for $W_3(x_1,x_2,x_3)$.} We have to consider the equalities,
\begin{align}
\hspace{-4mm}\label{eq:3pttotalder1}&\int \mathrm{d}X_1\mathrm{d}X_1^{\dagger} \mathrm{d}X_2\mathrm{d}X_2^{\dagger}\partial_{X_{1,ab}^{\dagger}}\left(\bigl[X_1^{\dagger}X_2^{\dagger}X_2 S_2^{k+1}\bigr]_{ab} \Tr(S_2^{q_1})\Tr(S_2^{q_2})e^{-N\Tr(X_1X_1^{\dagger})}e^{-N\Tr(X_2X_2^{\dagger})}\right) =0 \\
\hspace{-4mm}\label{eq:3pttotalder2}&\int \mathrm{d}X_1\mathrm{d}X_1^{\dagger} \mathrm{d}X_2\mathrm{d}X_2^{\dagger}\partial_{X_{2,ab}^{\dagger}}\left(\bigl[S_2^{k+1}X_1X_1^{\dagger}X_2^{\dagger}\bigr]_{ab} \Tr(S_2^{q_1})\Tr(S_2^{q_2})e^{-N\Tr(X_1X_1^{\dagger})}e^{-N\Tr(X_2X_2^{\dagger})}\right) =0\\    
\hspace{-4mm}\label{eq:3pttotalder-higher}&\int \mathrm{d}X_1\mathrm{d}X_1^{\dagger} \mathrm{d}X_2\mathrm{d}X_2^{\dagger}\partial_{X_{1,ab}^{\dagger}}\partial_{X_{2,bc}^{\dagger}}\left( \bigl[X_1^{\dagger}X_2^{\dagger}X_2 S_2^{k}X_1X_1^{\dagger}X_2^{\dagger}\bigr]_{ac}\Tr(S_2^{q_1})\Tr(S_2^{q_2})e^{-N\left(\Tr(X_1X_1^{\dagger})-\Tr(X_2X_2^{\dagger})\right)}\right)=0.
\end{align}
The inspection of these Schwinger-Dyson equations reveals that the only type of terms that we have not already faced are obtained when both derivatives $\partial_{X_{1,ab}^{\dagger}}$, $\partial_{X_{2,bc}^{\dagger}}$ distribute over the two traced monomials $\Tr(S_2^{q_1})$, $\Tr(S_2^{q_2})$. The distributed action of derivatives on the traced monomial leads to the term
\begin{equation}
    2q_1q_2\bE\left(\Tr\left(S_2^{q_1+q_2+k+1}\right)\right)=2q_1q_2m_{k+q_1+q_2}.
\end{equation}
The generating function of this term appearing in the corresponding loop equation will be 
\begin{multline}
 \sum_{k,q_1,q_2\ge 0}\frac{2q_1q_2m_{k+q_1+q_2}}{x_1^{k+1}x_2^{q_1+1}x_3^{q_2+1}}= \\
   \frac{2}{x_1}\frac{\partial^2}{\partial x_2\partial x_3}\left(\frac{(x_2-x_3)x_1x_2x_3W_1(x_1)-(x_1-x_3)x_1x_2x_3W_1(x_2)+(x_1-x_2)x_1x_2x_3W_1(x_3)}{\Delta(\{x_1,x_2,x_3\})} \right)
\end{multline}
where $\Delta(\{x_1,x_2,x_3\})=(x_3-x_2)(x_3-x_1)(x_2-x_1)$ is the Vandermonde determinant of  the family of variables $\{x_1,x_2,x_3\}$. The remaining terms of the loop equations can be inferred by realizing that for all terms involved in either \eqref{eq:3pttotalder1}, \eqref{eq:3pttotalder2}, \eqref{eq:3pttotalder-higher}, one of the two traced monomials plays a spectator role for the action of the derivatives. Consequently, one obtains the loop equation,
\begin{multline}
   0=\overline{W}_5(x_1,x_1,x_1,x_2,x_3)+\frac1{x_1}\partial_{x_1}\overline{W}_3(x_1,x_2,x_3)+\frac12\partial^2_{x_1}\overline{W}_3(x_1,x_2,x_3)-\frac{N^2}{x_1}\overline{W}_3(x_1,x_2,x_3)-N^2A_3(x_1,x_2,x_3)\\
   +\frac{1}{x_1^2}\partial_{x_2}\left( x_1x_2\frac{\overline W_3(x_1,x_1,x_3)-\overline W_3(x_1,x_2,x_3)}{x_1-x_2} \right)+\frac1{x_1^2}\partial_{x_2}\left( \frac{x_1x_2\overline W_3(x_1,x_1,x_3)-x_2^2\overline W_3(x_2,x_2,x_3)}{x_1-x_2}\right)\\
   +\frac{1}{x_1^2}\partial_{x_3}\left( x_1x_3\frac{\overline W_3(x_1,x_1,x_2)-\overline W_3(x_1,x_2,x_3)}{x_1-x_3} \right)+\frac1{x_1^2}\partial_{x_3}\left( \frac{x_1x_3\overline W_3(x_1,x_1,x_2)-x_3^2\overline W_3(x_3,x_3,x_2)}{x_1-x_3}\right)\\
   +\frac{2}{x_1^3}\frac{\partial^2}{\partial x_2\partial x_3}\left(\frac{(x_2-x_3)x_1x_2x_3W_1(x_1)-(x_1-x_3)x_1x_2x_3W_1(x_2)+(x_1-x_2)x_1x_2x_3W_1(x_3)}{\Delta(\{x_1,x_2,x_3\})} \right),
\end{multline}\\
where we have set $A_3(x_1,x_2,x_3)=-\frac{N}{x_1^2}\overline{W}_2(x_2,x_3)$. 

We now introduce some notations in order to shorten expressions. We denote
\begin{align}
    &\tilde{\cW}_{n+2}(x_1,x_1,x_1;x_2,\ldots,x_n)=\sum_{\mu\vdash [x_1,x_1,x_1]}\sum_{\bigsqcup_{i=1}^{|\mu|}J_i=\{x_2,\ldots,x_n\}}\prod_{\mu_i\in\mu}W_{|\mu_i|+|J_i|}(\mu_i,J_i)\\
    &\tilde{\cW}_{g,n+2}(x_1,x_1,x_1;x_2,\ldots,x_n)=\sum_{\mu\vdash [x_1,x_1,x_1]}\sum_{\substack{\bigsqcup_{i=1}^{|\mu|}J_i=\{x_2,\ldots,x_n\}\\\sum_{i=1}^{|\mu|}g_i=g+|\mu|-2}}\prod_{\mu_i\in\mu}W_{g_i,|\mu_i|+|J_i|}(\mu_i,J_i).
\end{align}
The notation $\mu \vdash [x_1,x_1,x_1]$ needs to be explained. The summation runs over the partitions $\mu$ of the list $[x_1,x_1,x_1]$ in the following sense. Firstly, in our notation the object $[x_a,x_b,x_c,\ldots]$ is a list of elements, that is an ordered multi-set. More concretely the order of appearance of the elements in the list is important and so for example the instances $[x_1,x_2,x_1,x_1,x_4]$, $[x_1,x_1,x_1,x_2,x_4]$ of lists are different (though they are the same multi-sets). We now come to explain what we mean by partitions of lists. A (denumerable\footnote{we will of course consider only the denumerable case since our lists are finite.}) list of elements can be represented as a set in the following way. We send a list to the set of pairs $\{(\textrm{element}, \textrm{position in the list})\}$. For instance, the list $[x_1,x_2,x_1,x_1,x_4]\mapsto\{(x_1,1),(x_2,2),(x_1,3),(x_1,4),(x_4,5)\}$ while the second list $[x_1,x_1,x_1,x_2,x_4]\mapsto \{(x_1,1),(x_1,2),(x_1,3),(x_2,4),(x_4,5)\}$ which are indeed two different sets. The partitions of the list $\mu$ are the partitions of the corresponding set of pairs $(\textrm{element}, \textrm{position in the list})$. However, note that the elements of the partitions forget about the position in the list and thanks to the symmetry of the functions $W_n$ functions should be seen as subsets of the corresponding multi-set. For instance, due to the fact that $\mu$ is really a partition of a list, the partition $\mu=\{\{x_1,x_1\},\{x_1\}\}$ with $\mu_1=\{x_1,x_1\}, \ \mu_2=\{x_1\}$ of the list $[x_1,x_1,x_1]$ appears three times in the sum. 

Some further notations are also required. The sum over $\bigsqcup_{i=1}^{|\mu|}J_i=\{x_2,\ldots,x_n\}$ means that we sum over the decompositions into $|\mu|$ (possibly empty) subsets $J_i$ of the set $\{x_2,\ldots,x_n\}$. For instance, in the case $n=3$, one can consider the term indexed by the partition $\mu=\{\{x_1,x_1\},\{x_1\}\}$ and the decomposition $J_1=\emptyset, \ J_2=\{x_2,x_3\}$, which correspond to a term of the form $W_2(x_1,x_1)W_3(x_1,x_2,x_3)$ in the sum. Note that these definitions are very similar to the ones appearing in \cite[Definition 4]{Bouchard-Eynard}. We also introduce the notation
\begin{equation}
    O_x=\frac1{x_1}\partial_{x_1}+\frac12\partial_{x_1}^2.
\end{equation}
Using these notations the corresponding equation for connected resolvents writes
\begin{align}\label{eq:3pt-loop-equation}
    0=&\tilde{\cW}_{5}(x_1,x_1,x_1;x_2,x_3)+O_xW_3(x_1,x_2,x_3)-\frac{N^2}{x_1}W_3(x_1,x_2,x_3)\nonumber\\
    &+\frac{2}{x_1^3}\frac{\partial^2}{\partial x_2\partial x_3}\left(\frac{(x_2-x_3)x_1x_2x_3W_1(x_1)-(x_1-x_3)x_1x_2x_3W_1(x_2)+(x_1-x_2)x_1x_2x_3W_1(x_3)}{\Delta(\{x_1,x_2,x_3\})} \right)\nonumber\displaybreak[1]\\
    &+\frac{1}{x_1^2}\partial_{x_2}\left( x_1x_2\left(\sum_{\substack{J\vdash[x_1,x_1,x_3]\\ J_i\neq\{x_3\}, \forall J_i}}\frac{\prod_{J_i\in J}W_{|J_i|}(J_i)}{x_1-x_2}- \sum_{\substack{J\vdash[x_1,x_2,x_3]\\ J_i\neq\{x_3\}, \forall J_i}}\frac{\prod_{J_i\in J}W_{|J_i|}(J_i)}{x_1-x_2}\right) \right)\nonumber\displaybreak[2]\\
    &+\frac{1}{x_1^2}\partial_{x_2}\left( x_1x_2\sum_{\substack{J\vdash[x_1,x_1,x_3]\\ J_i\neq\{x_3\}, \forall J_i}}\frac{\prod_{J_i\in J}W_{|J_i|}(J_i)}{x_1-x_2}- x_2^2\sum_{\substack{J\vdash[x_2,x_2,x_3]\\ J_i\neq\{x_3\}, \forall J_i}}\frac{\prod_{J_i\in J}W_{|J_i|}(J_i)}{x_1-x_2}\right)\nonumber\displaybreak[2]\\
    &+\frac{1}{x_1^2}\partial_{x_3}\left( x_1x_3\left(\sum_{\substack{J\vdash[x_1,x_1,x_2]\\ J_i\neq\{x_2\}, \forall J_i}}\frac{\prod_{J_i\in J}W_{|J_i|}(J_i)}{x_1-x_3}- \sum_{\substack{J\vdash[x_1,x_2,x_3]\\ J_i\neq\{x_2\}, \forall J_i}}\frac{\prod_{J_i\in J}W_{|J_i|}(J_i)}{x_1-x_3}\right) \right)\nonumber\displaybreak[2]\\
    &+\frac{1}{x_1^2}\partial_{x_3}\left( x_1x_3\sum_{\substack{J\vdash[x_1,x_1,x_2]\\ J_i\neq\{x_2\}, \forall J_i}}\frac{\prod_{J_i\in J}W_{|J_i|}(J_i)}{x_1-x_3}- x_3^2\sum_{\substack{J\vdash[x_3,x_3,x_2]\\ J_i\neq\{x_2\}, \forall J_i}}\frac{\prod_{J_i\in J}W_{|J_i|}(J_i)}{x_1-x_3}\right).\\
\end{align}
We can now extract the corresponding equation of order $g$ (that is the coefficient of $N^{-1-2g}$ in the expansion of \eqref{eq:3pt-loop-equation}). The corresponding family of equations on $W_{g,3}$ can then be solved recursively provided that we know the $W_{g',n'}$ of lower orders,
\begin{multline}\label{eq:3pt-loop-equation-expansion}
    0=\tilde{\cW}_{g,5}(x_1,x_1,x_1;x_2,x_3)+O_xW_{g-1,3}(x_1,x_2,x_3)
    -\frac{1}{x_1}W_{g,3}(x_1,x_2,x_3)\\
    +\frac{2}{x_1^3}\frac{\partial^2}{\partial x_2\partial x_3}\left(\frac{(x_2-x_3)x_1x_2x_3W_{g,1}(x_1)-(x_1-x_3)x_1x_2x_3W_{g,1}(x_2)+(x_1-x_2)x_1x_2x_3W_{g,1}(x_3)}{\Delta(\{x_1,x_2,x_3\})} \right)\displaybreak[1]\\
     +\frac{1}{x_1^2}\partial_{x_2}\left( x_1x_2\left(\sum_{\substack{J\vdash[x_1,x_1,x_3]\\ J_i\neq\{x_3\}, \forall J_i\\g=\sum_i g_i +4 -|J|}}\frac{\prod_{J_i\in J}W_{g_i,|J_i|}(J_i)}{x_1-x_2}- \sum_{\substack{J\vdash[x_1,x_2,x_3]\\ J_i\neq\{x_3\}, \forall J_i\\g=\sum_i g_i +4 -|J|}}\frac{\prod_{J_i\in J}W_{g_i,|J_i|}(J_i)}{x_1-x_2}\right) \right)\displaybreak[2]\\
     +\frac{1}{x_1^2}\partial_{x_2}\left( x_1x_2\sum_{\substack{J\vdash[x_1,x_1,x_3]\\ J_i\neq\{x_3\}, \forall J_i\\g=\sum_i g_i +4 -|J|}}\frac{\prod_{J_i\in J}W_{g_i,|J_i|}(J_i)}{x_1-x_2}- x_2^2\sum_{\substack{J\vdash[x_2,x_2,x_3]\\ J_i\neq\{x_3\}, \forall J_i\\g=\sum_i g_i +4 -|J|}}\frac{\prod_{J_i\in J}W_{g_i,|J_i|}(J_i)}{x_1-x_2}\right)\displaybreak[2]\\
     +\frac{1}{x_1^2}\partial_{x_3}\left( x_1x_3\left(\sum_{\substack{J\vdash[x_1,x_1,x_2]\\ J_i\neq\{x_2\}, \forall J_i\\g=\sum_i g_i +4 -|J|}}\frac{\prod_{J_i\in J}W_{g_i,|J_i|}(J_i)}{x_1-x_3}- \sum_{\substack{J\vdash[x_1,x_2,x_3]\\ J_i\neq\{x_2\}, \forall J_i\\g=\sum_i g_i +4 -|J|}}\frac{\prod_{J_i\in J}W_{g_i,|J_i|}(J_i)}{x_1-x_3}\right) \right)\displaybreak[2]\\
     +\frac{1}{x_1^2}\partial_{x_3}\left( x_1x_3\sum_{\substack{J\vdash[x_1,x_1,x_2]\\ J_i\neq\{x_2\}, \forall J_i\\g=\sum_i g_i +4 -|J|}}\frac{\prod_{J_i\in J}W_{g_i,|J_i|}(J_i)}{x_1-x_3}- x_3^2\sum_{\substack{J\vdash[x_3,x_3,x_2]\\ J_i\neq\{x_2\}, \forall J_i\\g=\sum_i g_i +4 -|J|}}\frac{\prod_{J_i\in J}W_{g_i,|J_i|}(J_i)}{x_1-x_3}\right).
\end{multline}
We now state in full generality the loop equations.\\

\noindent{\bf General loop equations.} We obtain the higher order loop equations in full generality by starting with Schwinger-Dyson equalities of the same type than \eqref{eq:3pttotalder1}, \eqref{eq:3pttotalder2}, \eqref{eq:3pttotalder-higher}, but we now insert more traces of monomials of the matrix $S_2$. Doing so we obtain more relations between  moments, and those relations can be translated into relations involving $W_n$ with higher values of $n$. As before, this first set of relations cannot be used to compute the $W_n$ as it does not close. To solve this problem we perform the $1/N$ expansion which leads to a closed set of equations on $W_{g,n}$. We display both the equations on $W_n$ and the equations on $W_{g,n}$ for $(g,n)$ such that $2g-2+n>0$. With $I_{ij}=\{x_1,\ldots,x_n\}\backslash\{x_i,x_j\}$,
\begin{multline}
    0=\tilde{\cW}_{n+2}(x_1,x_1,x_1;x_2,\ldots, x_n)+O_xW_{n}(x_1,\ldots,x_n)
    -\frac{N^2}{x_1}W_{n}(x_1,\ldots,x_n)\displaybreak[1]\\
    +\frac2{x_1^3}\sum_{\substack{2\le i<j\le n}}\frac{\partial^2}{\partial x_i\partial x_j}\left( \frac{(x_i-x_j)x_1x_ix_jW_{n-2}(I_{ij})-(x_1-x_j)x_1x_ix_jW_{n-2}(I_{1j})+(x_1-x_i)x_1x_ix_jW_{n-2}(I_{1i})}{\Delta(\{x_1,x_i,x_j\})}\right)\displaybreak[2]\\
    +\frac{1}{x_1^2}\sum_{i\in [\![ 2,n]\!]}\partial_{x_i}\left( x_1x_i\left(\sum_{\substack{J\vdash\{x_1,x_1\}\\ \bigsqcup_{k=1}^{|J|}K_k=\{x_2,\ldots,x_n\}\backslash\{x_i\}}}\hspace{-12mm}\frac{\prod_{J_l\in J}W_{|J_l|+|K_l|}(J_l,K_l)}{x_1-x_i}\, - \hspace{-6mm}\sum_{\substack{J\vdash\{x_1,x_i\}\\ \bigsqcup_{k=1}^{|J|}K_k=\{x_2,\ldots,x_n\}\backslash\{x_i\}}}\hspace{-12mm}\frac{\prod_{J_l\in J}W_{|J_l|+|K_l|}(J_l,K_l)}{x_1-x_i}\right) \right)\displaybreak[2]\\
     +\frac{1}{x_1^2}\sum_{i\in [\![ 2,n]\!]}\partial_{x_i}\left( x_1x_i\hspace{-6mm}\sum_{\substack{J\vdash\{x_1,x_1\}\\ \bigsqcup_{k=1}^{|J|}K_k=\{x_2,\ldots,x_n\}\backslash\{x_i\}}}\hspace{-12mm}\frac{\prod_{J_l\in J}W_{|J_l|+|K_l|}(J_l,K_l)}{x_1-x_i}\, - \hspace{2mm}x_i^2\hspace{-12mm}\sum_{\substack{J\vdash\{x_i,x_i\}\\ \bigsqcup_{k=1}^{|J|}K_k=\{x_2,\ldots,x_n\}\backslash\{x_i\}}}\hspace{-12mm}\frac{\prod_{J_l\in J}W_{|J_l|+|K_l|}(J_l,K_l)}{x_1-x_i}\right). 
\end{multline}
For the equations on $W_{g,n}$, write
\begin{multline}\label{eq:loop-eq-general-expanded}
    0=\tilde{\cW}_{g,n+2}(x_1,x_1,x_1;x_2,\ldots, x_n)+O_xW_{g-1,n}(x_1,\ldots,x_n)
    -\frac{1}{x_1}W_{g,n}(x_1,\ldots,x_n)\displaybreak[1]\\
    +\frac2{x_1^3}\sum_{\substack{2\le i<j\le n}}\frac{\partial^2}{\partial x_i\partial x_j}\left( \frac{(x_i-x_j)x_1x_ix_jW_{g,n-2}(I_{ij})-(x_1-x_j)x_1x_ix_jW_{g,n-2}(I_{1j})+(x_1-x_i)x_1x_ix_jW_{g,n-2}(I_{1i})}{\Delta(\{x_1,x_i,x_j\})}\right)\displaybreak[2]\\
    +\frac{1}{x_1^2}\sum_{i\in [\![ 2,n]\!]}\partial_{x_i}\left( x_1x_i\left(\sum_{\substack{J\vdash\{x_1,x_1\}\\ \bigsqcup_{k=1}^{|J|}K_k=\{x_2,\ldots,x_n\}\backslash\{x_i\}\\g=\sum_lg_l -|J|+2}}\hspace{-12mm}\frac{\prod_{J_l\in J}W_{g_l,|J_l|+|K_l|}(J_l,K_l)}{x_1-x_i}\, - \hspace{-6mm}\sum_{\substack{J\vdash\{x_1,x_i\}\\ \bigsqcup_{k=1}^{|J|}K_k=\{x_2,\ldots,x_n\}\backslash\{x_i\}\\g=\sum_lg_l -|J|+2}}\hspace{-12mm}\frac{\prod_{J_l\in J}W_{g_l,|J_l|+|K_l|}(J_l,K_l)}{x_1-x_i}\right) \right)\displaybreak[2]\\
     +\frac{1}{x_1^2}\sum_{i\in [\![ 2,n]\!]}\partial_{x_i}\left( x_1x_i\hspace{-6mm}\sum_{\substack{J\vdash\{x_1,x_1\}\\ \bigsqcup_{k=1}^{|J|}K_k=\{x_2,\ldots,x_n\}\backslash\{x_i\}\\g=\sum_lg_l -|J|+2}}\hspace{-12mm}\frac{\prod_{J_l\in J}W_{g_l,|J_l|+|K_l|}(J_l,K_l)}{x_1-x_i}\, - \hspace{2mm}x_i^2\hspace{-12mm}\sum_{\substack{J\vdash\{x_i,x_i\}\\ \bigsqcup_{k=1}^{|J|}K_k=\{x_2,\ldots,x_n\}\backslash\{x_i\}\\g=\sum_lg_l -|J|+2}}\hspace{-12mm}\frac{\prod_{J_l\in J}W_{g_l,|J_l|+|K_l|}(J_l,K_l)}{x_1-x_i}\right). 
\end{multline}
Using the family of equations \eqref{eq:loop-eq-general-expanded} one can recursively compute any $W_{g,n}$ knowing the initial conditions $W_{0,1}(x)$ and $W_{0,2}(x_1,x_2)$. Moreover, starting from these equations it should be possible to obtain a topological recursion like formula. Such a recursion formula certainly looks like the Bouchard-Eynard topological recursion formula introduced in \cite{Bouchard-and-al, Bouchard-Eynard}. Establishing such a formula strongly depends on the analytic properties of the $W_{g,n}$ as well as the geometric information contained in $W_{0,1}$ and $W_{0,2}$. Thus in the next section we try to make explicit some of these properties. We first focus on the geometry underlying the equation satisfied by $W_{0,1}$, and then describe the analytic properties of the higher order terms, by: 1. doing explicit computations and 2. studying the structure of the loop equations. A more detailed and systematic study of the analytical properties of the loop equations is postponed to further work on the product of $p$ rectangular Ginibre matrices.\\

\section{Spectral curve geometry}\label{sec:spectral-curve-geometry}
Before computing the first few solutions of the loop equations, we focus on studying the equation \eqref{eq:W01-equation} on $W_{0,1}$. Indeed, this equation defines an affine algebraic curve $\mathcal{C}$, called the spectral curve, where by affine algebraic curve we mean the locus of zero in $(x,y)\in \hat{\mathbb{C}}^2 = \left(\mathbb{C}\cup \{\infty\}\right)^2$ of the polynomial
\begin{equation}
    P(x,y)=x^2y^3-xy+1.
\end{equation}
This set of zeros of $P$ in $\mathbb{C}^2$ is generically a (complex) codimension $1$ subset of $\mathbb{C}^2$. In particular it can be given the structure of a Riemann surface. Computing the solutions $W_{0,1}(x)$ of \eqref{eq:W01-equation} gives a parametrization of the curve away from the ramification points. One of the goals of this section is to introduce a global, nicer parametrization called rational parametrization of the curve. Using this parametrization allows us to simplify the resulting expressions of the solutions. Indeed in the original $x$ variables, the solutions of \eqref{eq:W01-equation} are multi-valued. However one can fix that by promoting these solutions to meromorphic functions on the full affine curve defined by equation \eqref{eq:W01-equation}, the curve being the Riemann surface of $W_{0,1}(x)$.

\subsection{Basic properties of the curve}
\begin{figure}[t]
    \centering
    \includegraphics[scale=0.88]{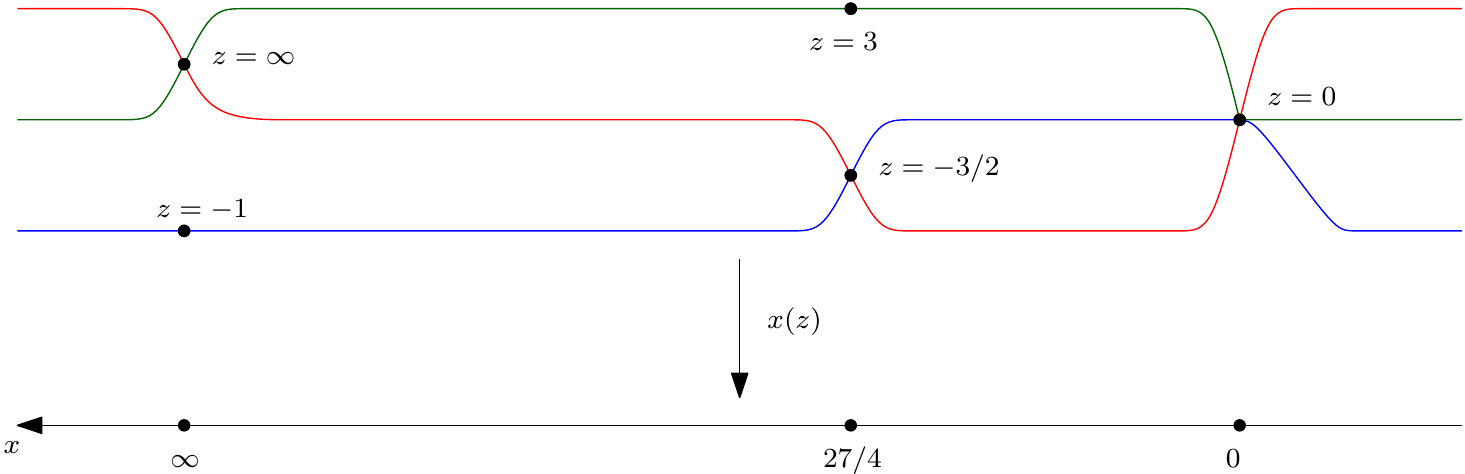}
    \caption{Ramification profile of the curve $\mathcal{C}$. We use colors to indicate permutations of sheets around ramification points.}
    \label{fig:ramif-profile}
\end{figure}

\begin{figure}[h]
    \centering
    \includegraphics[scale=0.9]{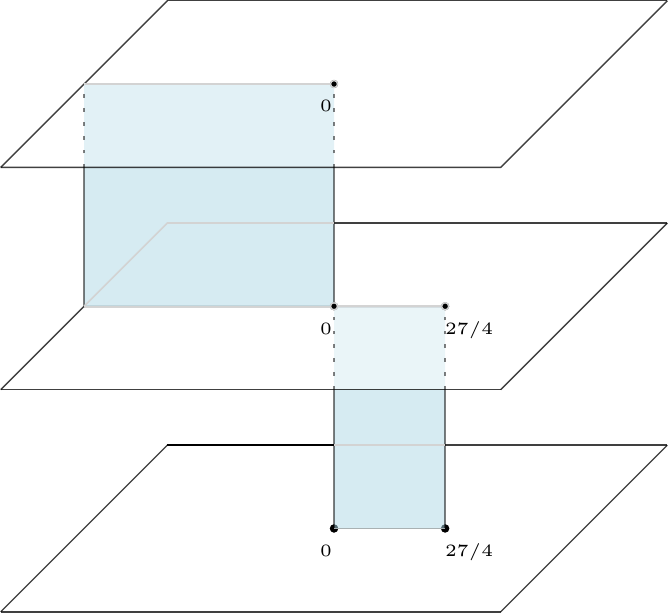}
    \caption{Cut structure of $W_{0,1}$.}
    \label{fig:cut-structure}
\end{figure}

There are two finite ramification points in the $x$-plane, one at $(x_{r_1},y_{r_1})=(27/4,2/9)$, which is a simple ramification point and one at $(x_{r_2},y_{r_2})=(0,\infty)$ which is a double ramification point. There is also one ramification point at infinity $x_{r_\infty}=\infty$ which is a simple ramification point. These ramifications are found from the condition that $P(x,y)=0$ and $\partial_yP(x,y)=0$. We display the ramification profile in Fig. \ref{fig:ramif-profile}.
The cut structure is readily described in \cite[Section 2.1 \& 2.2]{FLZ15}. It is pictured in Fig. \ref{fig:cut-structure}, where the lowest sheet of the figure corresponds to the \emph{physical} sheet  that is corresponding to the solution analytic at infinity, whose coefficients of the Laurent expansion are the moments of $S_2$. The other two sheets correspond to the two other solutions of \eqref{eq:W01-equation} that are not analytic at infinity. Indeed they have a simple ramification point at infinity.
From the Fig. \ref{fig:cut-structure} we can infer that the monodromy group is generated by the transposition $\tau_1 =(12)$ (obtained by going around $x_{r_1}$ in the physical sheet) and $\tau_2=(132)$ (going around $x_{r_2}$). These permutations are represented using colors on Fig. \ref{fig:ramif-profile}.

\begin{figure}
    \centering
    \includegraphics[scale=0.9]{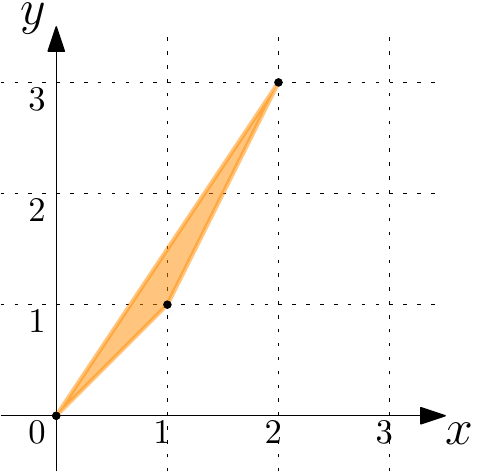}
    \caption{Newton polygon for the affine curve $x^2y^3-xy+1=0$. The number of $\mathbb{N}^2$ lattice points inside the polygon gives the generic genus of the curve. Here there is no points inside the polygon so that the generic genus is zero, which implies that the genus is zero.}
    \label{fig:newtonpolygon}
\end{figure}

The genus of the curve $\mathcal{C}$ can be obtained by considering the Newton polygon of the curve. The number of interior lattice points of the polygon drawn on Fig. \ref{fig:newtonpolygon} corresponds to the generic genus of the curve, that is the genus of the curve for generic enough coefficients of the polynomial $P$. However by fine tuning the coefficients of the polynomial one could in principle obtain a curve with smaller genus. The generic genus is the maximal genus the curve can have. In our case, $P(x,y)=x^2y^3-xy+1$, the number of lattice points in the Newton polygon is zero, thus the genus of the curve is zero. Since the genus of the curve is zero, there exists a rational parametrization. That is there exists two rational functions
\begin{align}
    &x:\hat{\mathbb{C}} \rightarrow \hat{\mathbb{C}}\\
    &y:\hat{\mathbb{C}} \rightarrow \hat{\mathbb{C}},
\end{align}
such that 
\begin{equation}
    x(z)^2y(z)^3-x(z)y(z)+1=0, \quad \forall z \in \hat{\mathbb{C}}.
\end{equation}
These two functions can be found by solving the following system on the coefficients of $Q_x(z),Q_y(z)$ and $P_x(z),P_y(z)$,
\begin{align}
   &Q_x(z) x(z)=P_x(z)\\
   &Q_y(z) y(z)=P_y(z)\\
   &x(z)^2y(z)^3-x(z)y(z)+1=0,
\end{align}
where $Q_x(z),Q_y(z)$ and $P_x(z),P_y(z)$ are set to be polynomials of degree high enough for a solution to exist. Then one obtains explicitly one possible parametrization 
\begin{equation}
    x(z)=\frac{P_x(z)}{Q_x(z)}=\frac{z^3}{1+z}, \quad y(z)=\frac{P_y(z)}{Q_y(z)}=-\frac{1+z}{z^2}.
\end{equation}
Note that from this point of view, $y(z)$ is the analytic continuation of $W_{0,1}(x(z))$. The function $x$ can be seen as a cover $x:\mathcal{C}\rightarrow \hat{\mathbb{C}}$ of generic degree $3$ (that is there are generically three values of $z$ corresponding to the same value of $x$). As such, the zeroes of $\mathrm{d}x$ corrrespond to the ramifications point of the cover. One can then check that $\mathrm{d}x=0$ at $z_{r_1}=0$ and $z_{r_2}=-3/2$, corresponding to the values $x(0)=0$ and $x(-3/2)=27/4$. One also notices that the zero of $\mathrm{d}x$ at $z=0$ is a double zero, thus confirming the fact that $x_{r_1}$ is a double ramification point. Finally since $x=27/4$ is a simple ramification point, there is another pre-image of $27/4$ in $z$ variable, that is we have $x(3)=27/4$. This leads to the ramification profile shown on Fig.\ref{fig:ramif-profile}. 
\subsection{Computation of $w_{0,1}$ and $w_{0,2}$}
Using this parametrization we compute the functions 
\begin{equation}\label{eq:wgn-def}
w_{g,n}(z_1,\ldots,z_n)=W_{g,n}(x(z_1),\ldots, x(z_n))\prod_{i=1}^n x'(z_i)+\frac{\delta_{g,0}\delta_{n,2}x'(z_1)x'(z_2)}{(x(z_1)-x(z_2))^2}.
\end{equation}
We also denote $\tilde{w}_{0,2}(z_1,z_2)=W_{0,2}(x(z_1),x(z_2))x'(z_1)x'(z_2)$. $w_{g,n}$ functions are meromorphic functions on $\mathcal{C}$, as such they are rational functions of their variables $z_i$. Consequently, they are much easier to manipulate than $W_{g,n}$ and their analytic properties are more transparent. For $w_{0,1}(z)$ we already know that $y(z) = W_{0,1}(x(z))$, thus 
\begin{equation}
    w_{0,1}(z)=y(z)x'(z)=-\frac{2z+3}{1+z}.
\end{equation}
\noindent The original functions $W_{g,n}$ can be recovered using the inverse function 
\begin{equation}
    z(x)=-xW_{0,1}(x)=_{\infty}-1-\frac{1}{x}-\frac{3}{x^2}-\frac{12}{x^3}-\frac{55}{x^4}+O\left(\frac{1}{x^5}\right).
\end{equation}
Indeed one has,
\begin{align}
    &W_{g,n}(x_1,x_2,\ldots,x_n)=\frac{w_{g,n}(z_1,z_2,\ldots, z_n)}{x'(z_1)x'(z_2)\ldots x'(z_n)}\Bigr\rvert_{z_i=z(x_i)} \textrm{ for } (g,n)\neq (0,2), \\
    &W_{0,2}(x_1,x_2)=\frac{\tilde w_{0,2}(z_1,z_2)}{x'(z_1)x'(z_2)}\Bigr\rvert_{z_1=z(x_1), z_2=z(x_2)}.
\end{align}
Note also that the corresponding coefficients of the expansion of $W_{g,n}$ at infinity, that is the $c^{[g]}_{k_1,\ldots, k_n}$, can be obtained by computing residues
\begin{equation}
  c^{[g]}_{k_1,\ldots, k_n}=\underset{\{x_i\rightarrow \infty\}}{\textrm{Res}}x_1^{k_1}\ldots x_n^{k_n}W_{g,n}(x_1,x_2,\ldots,x_n)= \underset{\{z_i\rightarrow -1\}}{\textrm{Res}}x(z_1)^{k_1}\ldots x(z_n)^{k_n}w_{g,n}(x(z_1),x(z_2),\ldots,x(z_n)).
\end{equation}
It is also true that the residue in $z$ variables can equivalently be computed at infinity. The passage from the $W_{g,n}$ to the $w_{g,n}$ functions takes into account the Jacobian of the change of variables.\\

For future convenience, we define
\begin{equation}
    \sigma(z)=\frac1{x(z)}(1-3x(z)y(z)^2),
\end{equation}
where $\sigma$ relates to $\partial_yP$ since $\sigma(z)=\frac1{x(z)^2}\partial_yP(x(z),y(z))$. So in particular $\sigma$ vanishes at the ramification point $(x_{r_1},y_{r_1})=(27/4,2/9)$ and $x(z)^2\sigma(z)$ has a zero of order $2$ at $(x_{r_2},y_{r_2})=(0,\infty)$. \\
\\

\noindent{\bf Expression of $\tilde{w}_{0,2}$.} We have after multiplying \eqref{eq:W02-equation} by $x'(z_1)x'(z_2)$ and performing a few additional manipulations
\begin{multline}\label{eq:w02-equation}
\sigma(z_1) \tilde{w}_{0,2}(z_1,z_2)=\frac{x'(z_1)}{x(z_1)^2}\partial_{z_2}\left(x(z_1) x(z_2)\frac{y(z_1)^2-y(z_1)y(z_2)}{x(z_1)-x(z_2)} \right) \\
+\frac{x'(z_1)}{x(z_1)^2}\partial_{z_2}\left( \frac{x(z_1)x(z_2)y(z_1)^2-x(z_2)^2y(z_2)^2}{x(z_1)-x(z_2)}\right).
\end{multline}
From this equation $\tilde{w}_{0,2}(z_1,z_2)$ can be computed in the variables $z_1,z_2$, so that one obtains
\begin{equation}\label{eq:explicit-tilde-w02}
    \tilde{w}_{0,2}(z_1,z_2)=\frac{z_2^2 z_1^2+2 (z_2 z_1^2+ z_2^2 z_1) + z_1^2 +z_2^2 +4 z_2 z_1}{(z_2 z_1^2+z_2^2
   z_1+z_1^2+z_2^2+z_2 z_1)^2}.
\end{equation}
From this expression we can recover the limiting cumulants of the product of traces,
\begin{equation}
c^{[0]}_{i,j}=\underset{z_1, z_2\rightarrow \infty}{\textrm{Res}}\, x(z_1)^ix(z_2)^j\tilde{w}_{0,2}(z_1,z_2).
\end{equation}\\
We provide the reader with the first few orders on Table \ref{tab:2pt-moments}. These numbers can be obtained easily \textit{via} symbolic computation softwares.
\begin{remark}\label{rem:guess_2pt-coeff}
Using a table of coefficients $c_{i,j}^{[0]}$ for $i,j$ running from $1$ to $20$ it is possible to make an experimental guess for the explicit form of these coefficients. This is
\begin{equation}
    c_{i,j}^{[0]}=\frac{2 i j }{3(i+j)}\binom{3i}{i}\binom{3j}{j}.
\end{equation}
In particular we have checked that these numbers satisfy the recurrence equation \eqref{eq:planar-bicumulant-equation} for the first few orders. It would be interesting to prove or disprove this guess \textit{via}, for instance, combinatorial means.
\end{remark}
\begin{table}[h]
\begin{center}
\begin{tabu}{l|c|c|c|c|c|c|c|c}
     \diag{.1em}{.8cm}{$j$}{$i$}& 1 & 2 & 3 & 4 & 5 & 6 & 7\\ \hline
      1 & 3   & 20   & 126  & 792    & 5005     & 31824     & 203490      \\ \hline
      2 & **  & 150  & 1008 & 6600   & 42900    & 278460    & 1808800     \\ \hline
      3 & **  & **   & 7056 & 47520  & 315315   & 2079168   & 13674528    \\ \hline
      4 & **  & **   & **   & 326700 & 2202200  & 14702688  & 97675200    \\ \hline
      5 & **  & **   & **   & **     & 15030015 & 101359440 & 678978300   \\ \hline
      6 & **  & **   & **   & **     & **       & 689244192 & 4649339520  \\ \hline
      7 & **  & **   & **   & **     & **       & **        & 31549089600 \\ \hline
\end{tabu}
\caption{Table of the first few cumulants $c^{[0]}_{i,j}=\lim_{N\rightarrow \infty}\bE\left(\Tr(S_2^i)\Tr(S_2^j)\right)-\frac{1}{N^2}\bE\bigl(\Tr(S_2^i)\bigr)\bE\bigl(\Tr(S_2^j)\bigr)$.\label{tab:2pt-moments}}
\end{center}
\end{table}

\noindent{\bf Universality for $w_{0,2}$.} In this paragraph we explain in detail and \textit{a posteriori}\footnote{Since they can already easily be inferred from the explicit result of equation \eqref{eq:explicit-tilde-w02}. } the analytic properties of $\tilde{w}_{0,2}$ and $w_{0,2}$. We first argue that $\tilde{w}_{0,2}$ does not have poles at the ramification points that is $z=-3/2,0$. We then consider the situation when $x(z_1)\rightarrow x(z_2)$. First starting from the above remark that $x(z)^2\sigma(z)=\partial_yP(x(z),y(z))$, we know that $x(z)^2\sigma(z)$ has a double zero at $z=0$ and a simple zero at $z=-3/2$, which makes it a source of poles as this factor appears in the denominator in front of the two terms of \eqref{eq:w02-equation-with-properfactors}, see below
\begin{multline}\label{eq:w02-equation-with-properfactors}
 \tilde{w}_{0,2}(z_1,z_2)=\frac{x'(z_1)}{x(z_1)^2\sigma(z_1)}\partial_{z_2}\left(x(z_1) x(z_2)\frac{y(z_1)^2-y(z_1)y(z_2)}{x(z_1)-x(z_2)} \right) \\
+\frac{x'(z_1)}{x(z_1)^2\sigma(z_1)}\partial_{z_2}\left( \frac{x(z_1)x(z_2)y(z_1)^2-x(z_2)^2y(z_2)^2}{x(z_1)-x(z_2)}\right).
\end{multline}
We start by focusing on poles at the simple ramification point $z=-3/2$. We remind ourselves that $\mathrm{d}x$ vanishes at the ramification points, and so $x'(z)$ has a simple zero at $z=-3/2$. Therefore $\frac{x'(z_1)}{x(z_1)^2\sigma(z_1)}$ is holomorphic at $z_1=-3/2$. Moreover, $x(z_1)$ and $y(z_1)$ are holomorphic at $z_1=-3/2$. As a consequence $\tilde{w}_{0,2}$ is holomorphic at $z=-3/2$ in both $z_1$ and $z_2$ (thanks to the symmetry $z_1 \leftrightarrow z_2$). 

We now come back to the ratio $\frac{x'(z_1)}{x(z_1)^2\sigma(z_1)}$ for $z_1=0$. A similar argument is valid at $z_1=0$. Indeed $x'(z_1)$ has a double zero at $z_1=0$ and this cancels the double zero of $x(z_1)^2\sigma(z_1)$ at $z_1=0$. In fact one can explicitly compute the ratio and find
\begin{equation}
    \frac{x'(z_1)}{x(z_1)^2\sigma(z_1)}=\frac1{1+z_1}
\end{equation}
which confirms our argument. $x(z)$ is holomorphic at $z=0$, but $y(z)$ is not, indeed it has a double pole at $z=0$. So the terms $x(z_1)x(z_2)y(z_1)^2$ could bring a simple pole at $z_1=0$. However, using the fact that $\tilde{w}_{0,2}(z_1,z_2)$ is symmetric in its arguments, if such a simple pole exists at $z_1=0$ then one should have a simple pole at $z_2=0$. Using the fact that $x(z_2)$ has a third order zero at $z_2=0$, and $y(z_2)$ has a double pole at $z_2=0$ one can show that $\tilde{w}_{0,2}(z_1,z_2)$ is holomorphic at $z_2=0$, therefore the apparent singularity at $z_1=0$ is a removable singularity. Consequently, we have just shown that $\tilde{w}_{0,2}(z_1,z_2)$ is holomorphic at the ramification points $z=-3/2, 0$ in both its variables.

Other possible singularities may occur at the singularities of $x(z)$ which possesses a simple pole at $z=-1$ and when $x(z_1)\rightarrow x(z_2)$. First note that 
\begin{equation}
    \frac{y(z_1)^2-y(z_1)y(z_2)}{x(z_1)-x(z_2)},
\end{equation}
has a double zero when $z_1\rightarrow -1$, thus 
\begin{equation}
\frac{x'(z_1)}{x(z_1)^2\sigma(z_1)}\partial_{z_2}\left(x(z_1) x(z_2)\frac{y(z_1)^2-y(z_1)y(z_2)}{x(z_1)-x(z_2)} \right)
\end{equation}
is holomorphic when $z_1\rightarrow -1$ since $\frac{x(z_1)x'(z_1)}{x(z_1)^2\sigma(z_1)}$ has a double pole at $z_1=-1$. A similar argument applies to the term 
\begin{equation}
    \frac{x'(z_1)}{x(z_1)^2\sigma(z_1)}\partial_{z_2}\left( \frac{x(z_1)x(z_2)y(z_1)^2-x(z_2)^2y(z_2)^2}{x(z_1)-x(z_2)}\right),
\end{equation}
thus showing that $\tilde{w}_{0,2}(z_1,z_2)$ is holomorphic at $z_1=-1$, and by symmetry at $z_2=-1$.\\

We are now left with the situation $x(z_1)\rightarrow x(z_2)$. A first possibility is $z_1\rightarrow z_2$. In this case both ratios
\begin{equation}\label{eq:2pt-ratio}
    \frac{y(z_1)^2-y(z_1)y(z_2)}{x(z_1)-x(z_2)}, \quad \frac{x(z_1)x(z_2)y(z_1)^2-x(z_2)^2y(z_2)^2}{x(z_1)-x(z_2)},
\end{equation}
are holomorphic since the denominators and numerators have simultaneous simple zeroes. So $\tilde{w}_{0,2}(z_1, z_2)$ is holomorphic when $z_1 \rightarrow z_2$. However, since $x(z)$ is a covering of degree three, there exists two (not globally defined) functions, $d_1(z), d_2(z)$ that leaves $x$ invariant, that is $x\circ d_i=x, \, i\in \{1,2\}$. These functions are the (non-trivial) solutions of the equation
\begin{equation}\label{eq:Deck-transf}
    \frac{d(z)^3}{1+d(z)}=\frac{z^3}{1+z}.
\end{equation}
This leads to the expressions
\begin{align}
    &d_1(z)=-\frac12\frac{z^2+z+z\sqrt{(z-3) (1+z)}}{1+z}, \\
    &d_2(z)=-\frac12\frac{z^2+z-z\sqrt{(z-3) (1+z)}}{1+z}.
\end{align}
One can check that $x(d_1(z))=x(d_2(z))=x(z)$. In order to understand the pole structure of $\tilde{w}_{0,2}(z_1, z_2)$, one also needs to know how does $y(z)$ changes when composed with one of the $d_i$. One has the simple identities for $i\in \{1,2\}$
\begin{equation}
    y(d_i(z))=\frac{d_i(z)}{z}y(z).
\end{equation}
Using these identities, one expects poles when $z_1 \rightarrow d_{1,2}(z_2)$. Indeed, in this limit the numerators of \eqref{eq:2pt-ratio} does not have zeroes anymore, while the denominators have simple zeroes. Thus $\tilde{w}_{0,2}(z_1, z_2)$ should have double poles when $z_1 \rightarrow d_{1,2}(z_2)$. This is indeed what we find by requiring that the denominator of \eqref{eq:explicit-tilde-w02} vanishes.

\begin{remark}
The functions $d_i$ have interesting properties. Indeed they permute the sheets of the covering $x:\mathcal{C}\rightarrow \hat{\mathbb{C}}$. Their behavior in a small neighborhood around a ramification point relates to the local deck transformation group of the cover. \\
Let us first focus on the double ramification point $z=0$. It is a fixed point of both $d_1$ and $d_2$ and around $z=0$, we have $d_1(z)\sim_0 e^{-\frac{2i\pi}{3}} z$ and $d_2(z)\sim_0 e^{\frac{2i\pi}{3}} z$ thus they are inverse of each other locally, and generate the cyclic group $\mathbb{Z}_3$. This cyclic group is the group generated by the permutation of the sheets $\tau_2=(132)$. This group is the local deck transformation group around the ramification point at $z=0$. \\
We now consider the behavior of $d_1, d_2$ at $z=-3/2$. In this case, only $d_1$ fixes $z=-3/2$, while $d_2(-3/2)=3,\, d_2(3)=-3/2$, that is $d_2$ exchanges the ramification point with the point above it (see Fig. \ref{fig:ramif-profile}). Note however that one has $d_1(3)=d_2(3)=-3/2$ as the two solutions $d_1,d_2$ of equation \eqref{eq:Deck-transf} merge at $z=3$ (as they also do at $z=1$). This merging has the following interpretation. At $z=-3/2$ two of the three sheets of the covering coincide. Therefore, there remains effectively only two sheets to be permuted, that is why $d_1$ fixes $z=-3/2$ while $d_2$ permutes $z=-3/2$ with $z=3$. The action of the local deck transformation group at $z=-3/2$ relates to the action of $d_1$ in a small neighborhood of $z=-3/2$. Since $d_1(-3/2+\epsilon)-d_1(-3/2)\sim_{0}-\epsilon$, $d_1$ locally generates the cyclic group $\mathbb{Z}_2$ corresponding to the group generated by the permutation $\tau_1=(12)$. Similar arguments can be used to describe the local deck transformation group at the ramification point $z=\infty$.
\end{remark}

We now come to the universality statement. Indeed, we expect that a slightly different object than $\tilde w_{0,2}(z_1,z_2)$ takes a universal form. This is the reason for the shift introduced in \eqref{eq:wgn-def}. The statement is that $w_{0,2}(z_1,z_2)$ should have a universal form, that is it should be the unique meromorphic function on the sphere with a double pole of order $2$ on the diagonal with coefficient $1$ and otherwise regular. Indeed if we compute $w_{0,2}(z_1,z_2)$ we obtain
\begin{equation}
    w_{0,2}(z_1,z_2)=\tilde w_{0,2}(z_1,z_2)+\frac{x'(z_1)x'(z_2)}{(x(z_1)-x(z_2))^2} = \frac{1}{(z_1-z_2)^2}.
\end{equation}
We find exactly the expected universal form for a genus zero spectral curve.\\

\noindent{\bf Comment on probabilistic interpretation of $W_{0,1}(x)$, $W_{1,1}(x)$ and $W_{0,2}(x_1,x_2)$.} As stated earlier, $W_1(x)$ is the Stieltjes transform of the eigenvalues density of the matrix $S_2$, that is 
\begin{equation}
    W_1(x)=\int_{-\infty}^{\infty}\mathrm{du} \frac{\rho_{1}(u)}{x-u}.
\end{equation}
In particular in the large $N$ limit we have that 
\begin{equation}
    W_{0,1}(x)=\int_{-\infty}^{\infty}\mathrm{du} \frac{\rho_{0,1}(u)}{x-u},
\end{equation}
and the computation of $W_{0,1}(x)$ uniquely determines $\rho_{0,1}(x)$. The same property is also true for the exact density, \textit{i.e.} $W_1(x)$ uniquely determines $\rho_1(x)$. This can be traced back to the Carlemann condition  \cite{akhiezer-moment-problem}. Indeed the Stieltjes transform $W_1(x)$, (resp. $W_{0,1}(x)$) contains the information on the whole moment sequence of $\rho_1(x)$ (resp. $\rho_{0,1}(x)$). The sequence of moments of both distributions can be shown to satisfy the Carlemann condition, and thus one expects that the knowledge of the Stieltjes transform is sufficient to reconstruct the densities $\rho_1(x)$, $\rho_{0,1}(x)$. However it is known \cite{forrester2006asymptotic} that in general the truncation of the $1/N$ expansion of the resolvent does not determine a unique truncated density. Indeed, there exists, \textit{a priori}, multiple densities truncated at order $p$, $\rho^{(p)}_1(x)=\sum_{g\ge 0}^pN^{-2g}\rho_{g,1}(x)$ with the same truncated resolvent
\begin{equation}
    \sum_{g\ge 0}^{p}N^{-2g}W_{g,1}(x) = \int_{-\infty}^{\infty}\mathrm{du} \frac{\rho^{(p)}(u)}{x-u}.
\end{equation}
That is the computation of the corrections to $W_{0,1}(x)$ only determines Stieltjes class of densities\footnote{Though this is not a rigorous justification, one can look at the truncated Carlemann criterion, for instance in the GUE case, and see that the Carlemann criterion is indeed not satisfied order-by-order in $1/N$. Only the large $N$ and the exact criterion are satisfied.}, often referred to as a \emph{smoothed} density. This is sufficient however to compute the corrections to the average $\bE(\phi(x))$ where $\phi(x)$ is any function analytic on the support of $\rho_{0,1}(x)$. In particular, our later computation of the first few corrections to the large $N$ resolvent does not determine corrections $\rho_{1,1}(x), \rho_{2,1}(x),\ldots$\\

The probabilistic interpretation of $W_{0,2}$ goes as follows. $W_{2}$ is the Stieltjes transform of the connected part of the eigenvalue correlation function
\begin{equation}
    W_{2}(x_1,x_2)=\int_{-\infty}^{\infty}\mathrm{d}u \mathrm{d}v \frac{\rho_2(u,v)}{(x_1-u)(x_2-v)},
\end{equation} 
and 
\begin{equation}
   \rho_2(x_1,x_2)=\bE\left(\sum_{i=1}^N\delta(x_1-\lambda_i)\sum_{j=1}^N\delta(x_2-\lambda_j)\right)- \rho_1(x_1)\rho_1(x_2),
\end{equation}
where the $\lambda_i$ are the eigenvalues of the matrix $S_2$. In the large $N$ limit, the centered random vector whose components are the traces of successive powers of the matrix $S_2$, $\left( \Tr(S_2^i)-\bE(\Tr(S_2^i))\right)_{i=1}^k$ converges to a normal random vector of zero mean and variance $\textrm{Var}_{m,n}$
\begin{equation}
    \textrm{Var}_{m,n}=c^{[0]}_{m,n}= \underset{z_1\rightarrow -1}{\textrm{Res}}\underset{z_2\rightarrow -1}{\textrm{Res}}x(z_1)^m x(z_2)^n w_{0,2}(z_1,z_2),
\end{equation}
where the normality of this \emph{centered} random vector at large $N$ follows from the fact that $W_n(x_1,\ldots,x_n)=O(1/N^{n-2})$, that is the higher cumulants of the limiting distribution of the family $\{\Tr(S_2^i)\}$ vanish at large $N$. This statement extends to the large $N$ limit of any linear statistics $A$ of the eigenvalues of the form 
\begin{equation}
    A=\sum_{i=1}^N a(\lambda_i),
\end{equation}
where $a$ is a sufficiently smooth function (analytic for instance), as we have
\begin{equation}
    \textrm{Var}(A)=\oint_{\Gamma}\oint_{\Gamma}\frac{dx_1 dx_2}{(2i\pi)^2} a(x_1)a(x_2) W_{0,2}(x_1,x_2),
\end{equation}
with $\Gamma$ a contour encircling the cut $(0,27/4]$ of $W_{0,1}(x)$.\\

\subsection{Computation of $w_{1,1}$ and higher correlation functions.} 
From these data one can access the first correction to the resolvent which allows in turn to access a first correction to the large $N$ density. The equation for $w_{1,1}(z)$ can be easily obtained from the equation \eqref{eq:1pt-NLO} on $W_{1,1}(x)$. It reads
\begin{equation}\label{eq:w11-z-variable-equation}
    w_{1,1}(z)=\frac{3x(z)^2}{x'(z)\partial_yP(x(z),y(z))}y(z)\tilde w_{0,2}(z,z) + \frac{x(z)^2}{\partial_yP(x(z),y(z))}\left(\partial_z y(z)-\frac{x''(z)}{2x'(z)^2}\partial_zy(z)+\frac1{2x'(z)}\partial^2_z y(z) \right).
\end{equation}
This leads to the result of the next paragraph.\\

\noindent{\bf Expression of $w_{1,1}(z)$ and analytic properties of \eqref{eq:w11-z-variable-equation}.} We obtain,
\begin{equation}
    w_{1,1}(z)=\frac{z^4+7 z^3+21 z^2+24 z+9}{z^2 (2 z+3)^4}.
\end{equation}
We notice that the poles are located at $z=0$ and $z=-3/2$, which are the zeroes of $\mathrm{d}x$. However, starting from \eqref{eq:w11-z-variable-equation} one can only infer that the poles of $w_{1,1}(z)$ can be located at $z=0, -3/2, -1$. Indeed, one can easily obtain from the analytic properties of $x(z), y(z)$ and $\tilde w_{0,2}(z,z)$ that the first term of the right hand side of \eqref{eq:w11-z-variable-equation} can have poles only at $z=0, -3/2$, and rule out singularities at $z=-1, \infty$. However when considering the derivatives term, that is the second term of equation \eqref{eq:w11-z-variable-equation}, one can not rule out poles at $z=-1$. The explicit computation shows that the coefficient of these poles is zero.
\begin{remark}
Note that we can also produce a guess for the coefficients $c^{[1]}_n$. We need however to prove our first guess of Remark \ref{rem:guess_2pt-coeff} for $c_{i,j}^{[0]}$ to be able to prove this guess using the Schwinger-Dyson equations. We provide our guess for purely informative purposes,
\begin{equation}
    c^{[1]}_n=\frac{(n-1)^2n}{6(3n-1)}\binom{3n}{n}.
\end{equation}
\end{remark}
~\\
\noindent{\bf Expression for higher correlations.} Using the loop equations \eqref{eq:loop-eq-general-expanded} we can compute any $n$-point resolvents recursively at any order. We illustrate this claim by providing the first few resolvents of higher order.\\
\noindent\textit{One point case.}
\begin{align}
    &w_{0,1}(z)=-\frac{2 z+3}{z+1}\\
   \label{eq:W11} &w_{1,1}(z)=\frac{z^4+7 z^3+21 z^2+24 z+9}{z^2 (2 z+3)^4}\\
   \label{eq:W21} &w_{2,1}(z)=\frac{9 z^9+153 z^8+1284 z^7+4227 z^6+7626 z^5+9246 z^4+8280 z^3+5220 z^2+1971 z+324}{z^3 (2 z+3)^{10}}.
\end{align}
\noindent\textit{Two points case.}
\begin{align}
    \label{eq:W02}&\tilde{w}_{0,2}(z_1,z_2)=\frac{z_2^2 z_1^2+2 (z_2 z_1^2+ z_2^2 z_1) + z_1^2 +z_2^2 +4 z_2 z_1}{(z_2 z_1^2+z_2^2
   z_1+z_1^2+z_2^2+z_2 z_1)^2}\\
   \label{eq:W12}&w_{1,2}(z_1,z_2)=\frac{pol(z_1,z_2)}{z_1^2 \left(2 z_1+3\right){}^6 z_2^2 \left(2 z_2+3\right){}^6},
\end{align}
with $pol(z_1,z_2)$ a symmetric polynomial of $z_1,z_2$ of degree $12$,
\begin{multline}
    pol(z_1,z_2)=128 z_2^6 z_1^6+1280 z_2^5 z_1^6+6144 z_2^4 z_1^6+12288 z_2^3 z_1^6+12480 z_2^2 z_1^6+6912 z_2 z_1^6+1728 z_1^6+1280 z_2^6 z_1^5+12800 z_2^5 z_1^5\\
    +55680 z_2^4 z_1^5+108672 z_2^3 z_1^5+111168
   z_2^2 z_1^5+62208 z_2 z_1^5+15552 z_1^5+6144 z_2^6 z_1^4+55680 z_2^5 z_1^4+215352 z_2^4 z_1^4+405000 z_2^3 z_1^4\\
   +414234 z_2^2 z_1^4+233280 z_2 z_1^4+58320 z_1^4+12288 z_2^6 z_1^3+108672
   z_2^5 z_1^3+405000 z_2^4 z_1^3+768312 z_2^3 z_1^3+809838 z_2^2 z_1^3+466560 z_2 z_1^3\\
   +116640 z_1^3+12480 z_2^6 z_1^2+111168 z_2^5 z_1^2+414234 z_2^4 z_1^2+809838 z_2^3 z_1^2+888165 z_2^2
   z_1^2+524880 z_2 z_1^2+131220 z_1^2+6912 z_2^6 z_1\\
   +62208 z_2^5 z_1+233280 z_2^4 z_1+466560 z_2^3 z_1+524880 z_2^2 z_1+314928 z_2 z_1+78732 z_1+1728 z_2^6+15552 z_2^5+58320 z_2^4\\
   +116640
   z_2^3+131220 z_2^2+78732 z_2+19683.
\end{multline}
\noindent\textit{Three points case.}
\begin{align}
    \label{eq:W03}w_{0,3}(z_1,z_2,z_3)=\frac{24}{\left(2 z_1+3\right){}^2 \left(2 z_2+3\right){}^2 \left(2 z_3+3\right){}^2}.
\end{align}
For all these computed $w_{g,n}$, $(g,n)\neq (0,1), (0,2)$ the poles are located at $z=0$ and $z=-3/2$. Therefore we can expect that the poles of $w_{g,n}$, for $2g-2+n>0$, are always located at $z=0$ and $z=-3/2$, however this remains to be proven.
\begin{remark}
The computed $w_{g,n}$ are rational functions of the $z_i$. We notice that the numerator of these rational functions seems to be a polynomial with positive integer coefficients. If this property is true for every $w_{g,n}$, it would be interesting to understand if these positive integers have an enumerative (combinatorics or geometry) meaning. 
\end{remark}
\section{Conclusion}

In this first paper on loop equations for matrix product ensembles, we have shown how to obtain loop equations for any resolvents for a random matrix defined as a product of two square complex Ginibre matrices without resorting to an eigenvalues or singular values reformulation of the problem. Indeed, the eigenvalues reformulation is yet to access these observable quantities. We used these loop equations to compute several terms of the expansion of the any resolvents $W_n$. In particular we accessed $W_{0,2}$, giving us information on the fluctuations of linear statistics, as well as the first correction $W_{1,1}$ to $W_{0,1}$. We expect a similar technique to apply to the more general case of the product of $p\ge 2$ rectangular Ginibre (complex or real) as well as to some other product ensembles, for instance the ensembles introduced in \cite{ForIps18} that are closely related to the Hermite Muttalib-Borodin ensemble. \\

Several questions are suggested by this work. The most straightforward one concerns the establishment of a topological recursion formula for the $w_{g,n}$. In the present case this topological recursion formula is certainly similar to the one devised in \cite{Bouchard-and-al, Bouchard-Eynard} by Bouchard and al. and Bouchard and Eynard. We postpone the construction of such formula to further works. Another interesting question oriented towards enumerative geometry concerns the application of the same technical means to the matrix model introduced by Ambj\o rn and Chekhov in \cite{A-C2014, A-C2018} which generates hypergeometric Hurwitz numbers. In these works the spectral curve is obtained, however this is done \textit{via} a matrix-chain approach that requires $p-1$ of the $p$ matrices to be invertible, thus ruling out the fully general case of rectangular matrices. We hope this fully general case can be tackled using our \emph{higher} derivatives technique.\\  

Yet another related question is the following. Free probability provides us with tools to determine the equation satisfied by the large $N$ limit of the resolvent of a product of matrices knowing the large $N$ limit of the resolvents of the members of the product. These tools have been generalized to some extent to the $2$-point resolvent in the works of Collins and al. \cite{collins2007second} in order to more systematically access the fluctuations of linear statistics. One question is then the following. Can we devise similar tools that would allow to construct the full set of loop equations for a product matrix knowing the loop equations satisfied by the member of the product (or, more realistically, the large $N$ sector of the loop equations)? \\

Finally, the loop equations can be interpreted as Tutte equations \cite{countingsurfaces, tutte1962, tutte1968}. The loop equations described in this paper can also be interpreted combinatorially, and it would be interesting to understand the more general case of maps with an arbitrary number of black vertices in such a combinatorial setting. Moreover, one would also like to understand if it is possible to merge two sets of Tutte equations together for two independent sets of maps with one type of edge in order to obtain Tutte equations for maps with two types of edges. The combinatorial interpretation of the free multiplicative convolution described in \cite[section 3.3]{DLN} may be a useful starting point.
\bibliographystyle{alpha}
\bibliography{Typical-RMT-biblio}
\end{document}